\documentclass[aps, prc, twocolumn, superscriptaddress, nofootinbib, 10pt]{revtex4-2}
\usepackage{graphicx}  
\usepackage{booktabs}  
\usepackage[colorlinks=true,linkcolor=blue,citecolor=blue,urlcolor=blue]{hyperref}
\usepackage{amsmath}   
\usepackage{dcolumn} 
\usepackage{bm}   
\usepackage{float}
\usepackage[section]{placeins}

\usepackage{dblfloatfix}

\begin{document}

\title{Hierarchical Neural Filtering of Nuclear Mass Residuals and Spectral Signatures of Quantum Chaos}
\author{Jaskirat Singh}
\affiliation{MS23120, Department of Physical
Sciences, IISER Mohali, India}

\author{Chong Qi}
\email{chongq@kth.se}
\affiliation{Department of Physics, Royal
Institute of Technology (KTH), SE-106 91
Stockholm, Sweden}
\vspace{1cm}

\begin{abstract}
In complex quantum many-body systems such as atomic nuclei, the interplay between regular collective
motion and irregular intrinsic dynamics gives rise to fluctuations that cannot be fully captured by existing global theoretical models. Nuclear mass, which exhibits smooth trends across the nuclear chart together
with localized deviations, provides a sensitive observable for investigating such irregular
dynamics. In this work, we employ a variety of neural network architectures, which serve as controlled nonlinear filters within a Hierarchical Residual Decomposition framework to progressively extract and suppress the chaotic many-body signature (characterized by $1/f$ spectral correlations) in nuclear mass residuals. The resulting Physics-Informed Neural Ensemble (PINE) model combines multiple mass models and neural network architectures, enabling a systematic suppression of coherent and chaotic components, after which the remaining fluctuations are analyzed using Fourier-based spectral diagnostics across different mass regions.
Our results show that hierarchical neural residual learning efficiently removes the dominant low-frequency correlations and suppresses the quantum-chaotic spectral rigidity, driving the residuals toward the uncorrelated white-noise limit. This systematic suppression provides a quantitative diagnostic of the underlying scale-dependent complexity and many-body correlation structure of nuclear mass deviations.
\end{abstract}

\maketitle

\section{Introduction}
The study of nuclear structure physics has long been dominated by the emergence of regular and highly predictable phenomena such as the nuclear shell structure and collective excitations~\cite{BohrMottelsonI}. On the other hand, the nucleus is a typical complex many-body system where regular motion frequently coexists with irregular, unpredictable dynamics. This behavior arises from the intricate interplay of nonlinearity and strong interaction coupling in a high-dimensional dynamical regime. In classical mechanics, deterministic chaos is characterized by a high sensitivity to initial conditions and diverging phase-space trajectories. In the quantum many-body domain, signatures of chaos appear through the statistical properties of observables, spectral correlations, and response functions~\cite{Guhr1998,RevModPhys.81.539,Gomez2011-oq,BORGONOVI20161,Weidenmuller2025-ry,Mondal2026,PhysRevLett.126.150601,Yoshimura2024-rl,Xiang2024-pg,PhysRevE.111.054110,PhysRevLett.112.193003,PhysRevLett.131.166401,Pausch_2025}. Quantum chaos is not synonymous with randomness~\cite{Gomez2017-il}. Instead, it reflects structured complexity emerging from deterministic microscopic dynamics. Atomic nuclei serve as a prime example of such complex quantum systems. The chaotic behavior in nuclei is generally attributed to residual many-body correlations that lie beyond mean-field approximations and are absent in exactly integrable systems. These effects make the system sensitive to small changes in particle number and deformation, particularly in regions where multiple configurations compete. In fact, studies of this kind of complexity started even before the modern shell model was developed, originating with Wigner's pioneering work on spectral statistics~\cite{Wigner1951}, where Random Matrix Theory~\cite{Hsu1939,Wigner1967,Mehta2004} was introduced to describe fluctuations in highly excited nuclear spectra. It was subsequently demonstrated that the statistical distribution of energy levels provides a primary diagnostic of the system's underlying chaotic dynamics~\cite{Dyson1962I,Dyson1962II,Dyson1962III,Mehta1960,Bohigas1984,Relano2002,PhysRevLett.93.244101,PhysRevLett.118.204101,Casal2021-jk,PhysRevC.102.044301,Khatoni2021-gy,Mourik2018-ij,PhysRevLett.132.220401}.

Because of these complex many-body effects, nuclear observables that depend on detailed structural correlations, such as binding energies and excitation spectra, often exhibit irregular fluctuations superimposed upon otherwise smooth global trends. When nuclear observables are analyzed as discrete sequences across the nuclear chart, chaotic or irregular components manifest as short-range correlations and broadband spectral features, clearly distinct from smooth collective behavior. Such systems may reside in a regime intermediate between regular motion and fully developed chaos. Identifying these contributions requires diagnostic tools capable of disentangling coherent large-scale structure from locally fluctuating components, thereby helping to elucidate the structure, origin, and persistence of irregular dynamics that emerge naturally from the many-body character of the nucleus~\cite{Aberg2002}.

Nuclear binding energies, in particular, exhibit a pronounced regularity that can be described with high accuracy within macroscopic--microscopic frameworks. A detailed analysis of nuclear mass fluctuations in Ref.~\cite{PhysRevLett.94.102501}, for example, examined the statistical properties of residuals across different global mass models.
A central question is whether the remaining fluctuations can be eliminated through further model refinement or whether they reflect intrinsic quantum-chaotic effects. Despite significant theoretical advances, small but persistent residuals remain between experimental data and all existing nuclear mass models, potentially signaling underlying chaotic dynamics~\cite{Hirsch2004-sl,Hirsch2005-rr,PhysRevLett.96.042502,VELAZQUEZ2005134,MOLINARI200648,PhysRevLett.88.092502,PhysRevLett.94.102501}. From this perspective, residual fluctuations should not be regarded merely as model deficiencies or statistical artifacts. They may encode physically meaningful information about irreducible complexity inherent in the nuclear many-body problem~\cite{PhysRevLett.94.102501,Qi2026_Strutinsky,Storbacka2026}.

Considerable progress has been made in nuclear mass modeling through the incorporation of machine-learning and neural-network (NN) techniques, with predictive accuracies approaching or even surpassing the 100 keV level~\cite{Gernoth1993,PhysRevC.93.014311,PhysRevC.96.044308,Tian2025,PhysRevResearch.2.043363,Huang2025,universe7050131,NIU2018759,PhysRevC.109.064322,LI2024138385,PhysRevC.106.014305,PhysRevC.106.L021301,PhysRevC.111.024316}. These developments demonstrate that nonlinear data-driven methods can effectively capture systematic trends not fully described by traditional theoretical frameworks~\cite{Gernoth1993,PhysRevC.93.014311,PhysRevC.96.044308,PhysRevC.106.L021301,PhysRevC.106.014305,PhysRevC.109.064322,LI2024138385,universe7050131,Huang2025}. Despite these advances, a fundamental question remains: do residual discrepancies between experiment and theory vanish with continued model refinement, or does a persistent component remain that is insensitive to improvements in the global description? If such a component survives across different theoretical frameworks and filtering procedures, it may signal intrinsic many-body dynamics rather than deficiencies of a particular mass model~\cite{PhysRevLett.88.092502,Hirsch2004-sl,Hirsch2005-rr,PhysRevLett.96.042502,MOLINARI200648}. Addressing this issue requires a method capable of isolating smooth, learnable structure while preserving any residual complexity that lacks coherent correlations. Rapid progress has been made in data-driven and physics-informed approaches to nuclear mass prediction, including convolutional architectures, recurrent neural frameworks, kernel-based methods, and feature-guided learning strategies~\cite{PhysRevC.109.034318,PhysRevC.111.014325,rzjx-9zz1,Jalili2025,PhysRevC.111.034305}.

In this work, NNs are employed not merely as tools for maximizing predictive accuracy, but rather as controlled nonlinear filters applied to nuclear mass residuals. We construct a hybrid mass model in which an NN is trained specifically on the residuals:
\begin{equation}
\Delta M(Z,N)=M(Z,N)_{\mathrm{exp}}-M(Z,N)_{\mathrm{model}}
\label{eq:deltam}
\end{equation}
Rather than learning the full experimental mass surface $M_{\mathrm{exp}}(N,Z)$ directly, as is often done in purely data-driven approaches~\cite{PhysRevC.93.014311,PhysRevC.96.044308,PhysRevC.106.014305,PhysRevC.109.064322,LI2024138385}, the NN is restricted to learning only the structure not captured by the underlying physical model. In this sense, the global nuclear mass model provides the dominant coarse control of the mass surface, while the NN acts as a finer corrective mechanism that adjusts the remaining localized discrepancies. Improving extrapolation reliability and generalization capability remains a challenge in NN-based nuclear mass modeling~\cite{Zhao2022Generalization,mcxf-d32x,PhysRevC.111.024306,Sundberg2025Criticality}. Here, the network is trained with progressively increasing capacity to systematically resolve and suppress the chaotic correlations present in the residuals. The resolution threshold required to eliminate these spectral signatures serves as a quantitative physical diagnostic of the underlying fluctuation complexity~\cite{PhysRevLett.96.042502,MOLINARI200648}.

To study this behavior, we use a hierarchy of residual learners based on both fully connected feedforward neural networks (FFNNs) and mixture-of-experts (MoE) architectures~\cite{Hornik1989,Cybenko1989,Goodfellow2016,Jacobs1991,Jordan1994}. This constitutes a Hierarchical Residual Decomposition (HRD) framework that progressively extracts learnable structure from the residuals. The resulting Physics-Informed Neural Ensemble (PINE) model combines outputs from multiple network architectures to evaluate and isolate these correlations. In this approach, the FFNN components provide smooth global function approximation across the nuclear chart, while the MoE structures introduce adaptive specialization by allowing different expert subnetworks to focus on distinct residual regimes and local fluctuation patterns~\cite{Jacobs1991,Jordan1994}. Since neither architecture contains explicit sequential memory or imposed ordering dependence, the extracted correlations arise intrinsically from the nuclear input space itself rather than from artificial sequence learning. The combined framework therefore acts as a physics-guided spectral filtering mechanism, whose capacity to suppress chaotic many-body signatures ($1/f$ correlations) and drive the residual field toward the white-noise limit is quantified in the results below.

By combining NN-based filtering with spectral analysis, we show that the chaotic many-body signature ($1/f$ correlations) can be systematically suppressed by the high-capacity neural network hierarchy. We present this filtering process as a multi-scale resolution analysis, where we explore the transition of the residuals into a spectrally flat white-noise regime as a function of network capacity and regularization strength. This allows us to characterize the scale-dependent transition between coherent physical correlations and purely stochastic fluctuations.

\section{Methodology}
\subsection{Nuclear mass models and the nature of their residuals}
Nuclear mass models aim to describe binding energies across the nuclear chart by combining a limited set of physical ingredients that capture the dominant collective and single-particle contributions to nuclear structure~\cite{FRDM2012,wu2015global,wang2014surface,HFB24,Batail2025,DZ10,qi2015theoretical}. Although the formulations of global mass models differ substantially, most of them separate the smooth macroscopic behavior associated with the bulk properties of nuclear matter from microscopic corrections arising from shell structure, pairing correlations, and deformation effects. Macroscopic--microscopic approaches implement this philosophy explicitly by constructing a smooth liquid-drop-like energy functional supplemented by corrections obtained from an underlying single-particle potential~\cite{FRDM2012,wu2015global,wang2014surface,Qi2026_Strutinsky}. In contrast, self-consistent mean-field mass models derive nuclear masses from effective nucleon--nucleon interactions, treating deformation and pairing on the same footing within an energy density functional framework, refined by empirical corrections~\cite{HFB24,Batail2025}. While these approaches differ substantially in their physical assumptions and computational implementation, they all seek to reproduce the same experimental mass surface using a finite and necessarily incomplete representation of the underlying many-body dynamics~\cite{Lunney2003,FRDM2012,HFB24,DZ10,qi2015theoretical}. Because of these limitations, no global mass model can fully account for all correlations present in experimental data~\cite{PhysRevLett.88.092502,PhysRevLett.96.042502,MOLINARI200648}. The discrepancies between calculated and measured binding energies, commonly expressed as mass residuals, $\Delta M$, therefore contain valuable information about the physics not captured by the model~\cite{Hirsch2004-sl,Hirsch2005-rr,PhysRevLett.94.102501}. These residuals are not expected to be purely random. Smooth, long-range deviations may reflect systematic flaws in the macroscopic or mean-field description, while localized fluctuations may originate from missing correlations or chaotic many-body dynamics~\cite{Aberg2002,PhysRevLett.88.092502,PhysRevLett.96.042502}.

From a statistical perspective, mass residuals may be viewed as discrete signals defined over the nuclear chart. Their structure reveals the degree of coherence and complexity remaining after accounting for known physics~\cite{Relano2002,PhysRevLett.93.244101,PhysRevLett.118.204101}. Models that emphasize global smoothness may leave behind residuals that are small in magnitude but correlated over extended regions, whereas models incorporating more local correlations may produce residuals that are less coherent but more fragmented~\cite{PhysRevLett.94.102501,NIU2018759}. Understanding these differences is essential not only for assessing model performance, but also for identifying signatures of intrinsic many-body complexity that persist regardless of the modeling framework~\cite{MOLINARI200648,Gomez2011-oq,RevModPhys.81.539}.

In this context, nuclear mass models provide a controlled setting where the interplay between regular collective behavior and irregular many-body dynamics can be systematically explored~\cite{BohrMottelsonI,BohrMottelsonII,Bender2003,doi:10.1142/S021830132630002X}. By comparing the residual structures associated with different theoretical approaches, the influence of various physical assumptions on the nature, scale, and persistence of irregular components in nuclear binding energies can be investigated~\cite{FRDM2012,HFB24,PhysRevLett.94.102501}.

\subsection{Spectral decomposition of residuals and frequency analysis}

When nuclear observables such as binding energies are written as discrete sequences, their fluctuations can be studied using Fourier and spectral methods~\cite{Relano2002,PhysRevLett.93.244101,PhysRevLett.118.204101}. The Fourier transform decomposes the signal into components characterized by distinct frequencies and correlation lengths. Smoothly varying, large-scale structures manifest primarily in the low-frequency domain, whereas localized, rapid fluctuations contribute predominantly to the higher-frequency regime.

For nuclear mass residuals, the low-frequency domain is associated with smooth, collective trends extending over broad regions of the nuclear chart~\cite{PhysRevLett.94.102501,NIU2018759,Hirsch2004-sl}, whereas the high-frequency regime reflects localized variations and is highly sensitive to changes in nucleon numbers. Analyzing these spectral regions allows one to quantify the persistence of correlated structures after the subtraction of smooth background behaviors. A prominent low-frequency component indicates the survival of long-range correlations in the residuals; conversely, a flatter spectrum indicates weaker, increasingly fragmented fluctuations~\cite{Bohigas1984,Guhr1998,RevModPhys.81.539,Gomez2011-oq}.

The physical interpretation of the high-frequency regime depends on the nature of the fluctuations inherent in the system. Purely stochastic noise yields a flat power spectrum, where spectral power is uniformly distributed across all frequencies~\cite{Relano2002}. In contrast, structured many-body dynamics can generate spectra characterized by nontrivial frequency dependencies and approximate power-law scaling over broad intervals~\cite{Bohigas1984}, indicating that the underlying correlations persist across multiple scales rather than behaving as purely random fluctuations.

Spectral analysis thus provides a robust diagnostic framework to compare the residual structures obtained from different nuclear mass models~\cite{Relano2002,PhysRevLett.93.244101,PhysRevLett.118.204101}. Models that retain coherent large-scale structures within their residual fields exhibit enhanced low-frequency spectral power. Conversely, models that incorporate local correlations yield flatter spectra with diminished long-range structure~\cite{PhysRevLett.94.102501,Hirsch2004-sl,PhysRevLett.96.042502}. Variations in the spectral slope therefore provide a direct monitor of the progressive extraction of correlated structures from the residuals.

\subsubsection*{Boustrophedon ordering}
To probe correlations in nuclear mass residuals, the two-dimensional $(Z,N)$ nuclear chart is mapped onto one-dimensional sequences using two complementary boustrophedon orderings. In the first construction, nuclei are grouped according to increasing mass number $A$ and arranged sequentially within each isobaric chain using an alternating zigzag ordering in the neutron--proton asymmetry coordinate $(N-Z)$~\cite{PhysRevLett.94.102501}. Specifically, neighboring isobaric chains are traversed in opposite $(N-Z)$ directions in order to preserve continuity of the residual sequence while retaining correlations associated with shell evolution, deformation, pairing, and isospin-dependent structures across neighboring nuclei. This construction therefore provides a generic projection of the residual surface in which different physical mechanisms remain intermixed.

In the second construction, nuclei are reorganized according to their distances from the nearest proton and neutron magic numbers~\cite{BohrMottelsonI}. The shell distances are defined as
\begin{equation}
d_Z = \min\{|Z-Z_m|\},
\qquad
d_N = \min\{|N-N_m|\}
\label{eq:magic_distance}
\end{equation}
where $Z_m$ and $N_m$ denote the proton and neutron magic numbers,
\[
2,\ 8,\ 20,\ 28,\ 50,\ 82,\ 126.
\]
A combined shell-distance coordinate is then introduced as
\begin{equation}
d = d_Z + d_N
\label{eq:total_distance}
\end{equation}
The nuclei are subsequently separated into three shell-distance sectors. The first group ($g_1$) contains nuclei satisfying $d<5$, $d_Z\leq2$ and $d_N\leq2$, corresponding to nuclei located very close to both proton and neutron shell closures. The third group ($g_3$) contains nuclei with $d\geq6$, representing strongly deformation-dominated regions far from shell closures. The \emph{intermediate} group ($g_2$) consists of the remaining nuclei with $d<6$, thereby describing transitional shell-to-deformation regions. Within each group, nuclei are first arranged according to increasing values of $d$, and nuclei sharing the same $d$ are traversed according to $(N-Z)$ using an alternating boustrophedon ordering in which successive shell layers are scanned in opposite directions. Residual degeneracy is inherited from the original chart ordering, effectively preserving increasing mass-number continuity. The final one-dimensional sequence is obtained through the concatenation $(g_1,g_2,g_3)$, progressing systematically from shell-stabilized to deformation-dominated regions~\cite{PhysRevLett.94.102501,Hirsch2005-rr}.

Comparing the spectral properties from these two complementary orderings allows shell-driven patterns to be distinguished from other sources of residual correlations. It simultaneously tests the robustness of the inferred spectral behavior against changes in the geometric projection of the nuclear chart.

\subsubsection*{One-dimensional Fourier spectral analysis}

After the boustrophedon ordering, the nuclear mass residuals form a one-dimensional discrete sequence
\[
\Delta M(j),
\]
where $j$ denotes the boustrophedon index.

To suppress artificial discontinuities at the sequence boundaries during the Fourier analysis, the residual sequence is symmetrically mirrored only for the transform calculation,
\[
\Delta M_{\mathrm{mirror}}
=
\{
\Delta M_1,\ldots,\Delta M_N,
\Delta M_N,\ldots,\Delta M_1
\}.
\]
The mirrored residual sequence is then centered and normalized according to
\begin{equation}
\widetilde{\Delta M}(j)
=
\frac{
\Delta M(j)-\langle \Delta M\rangle
}{
\sigma_{\mathrm{rms}}
}
\label{eq:normalized_residual}
\end{equation}
where
\begin{equation}
\sigma_{\mathrm{rms}}
=
\sqrt{
\frac{1}{N}
\sum_{j=1}^{N}
\left(
\Delta M(j)-\langle \Delta M\rangle
\right)^2 }
\label{eq:rms}
\end{equation}
The discrete Fourier transform of the normalized sequence is then defined as
\begin{equation}
F(k)
=
\frac{1}{\sqrt{N}}
\sum_{j=1}^{N}
\widetilde{\Delta M}(j)
\, e^{-2\pi i k j/N}
\label{eq:dft}
\end{equation}
with corresponding power spectrum
\begin{equation}
P(k)=|F(k)|^2
\label{eq:power_spectrum}
\end{equation}
The Fourier power spectrum provides a decomposition of the residual fluctuations according to their characteristic correlation scales. Low-frequency components correspond to slowly varying collective structures extending over broad regions of the nuclear chart, whereas high-frequency components reflect localized fluctuations and rapidly varying correlations. By examining the distribution of spectral power across frequencies, one can distinguish coherent long-range behavior from fragmented or short-range residual structure~\cite{Relano2002,PhysRevLett.93.244101,Bohigas1984}.

The spectral properties are analyzed through the power-law behavior of the Fourier spectrum in the logarithmic representation,
\begin{equation}
P(k) \sim k^{-\alpha},
\end{equation}
where the scaling exponent $\alpha$ (yielding a spectral slope of $-\alpha$ in a logarithmic representation) quantifies the degree of correlation in the residual sequence. Large values of $\alpha$ (steep negative slopes) indicate dominant low-frequency coherence associated with smooth collective trends, whereas values of $\alpha$ approaching zero correspond to a spectrally flat regime characteristic of weakly correlated or fragmented fluctuations. Intermediate slopes characterize residual structures that are neither fully coherent nor purely stochastic, retaining localized correlations over restricted regions of the nuclear chart~\cite{Relano2002,PhysRevLett.93.244101,PhysRevLett.118.204101}.

Within this framework, the evolution of the spectral slopes before and after NN filtering provides a direct measure of how efficiently correlated structures are removed from the residuals. A substantial reduction of low-frequency power indicates suppression of smooth systematic trends, while the persistence of nonvanishing spectral slopes after filtering signals the survival of residual correlations that are not fully captured within the adopted modeling framework~\cite{Aberg2002,PhysRevLett.96.042502,MOLINARI200648,Richardson2025DNA,PhysRevC.111.044317,Sundberg2025Criticality}.

\subsubsection*{Two-dimensional Fourier analysis}

Although the one-dimensional boustrophedon orderings preserve important local correlations, any mapping of the nuclear chart onto a single sequence inevitably mixes structures originating from different directions in the $(Z,N)$ plane. To examine the residual fluctuations without imposing a specific sequential ordering, a complementary two-dimensional Fourier analysis is performed directly on the discrete nuclear lattice~\cite{NIU2018759,PhysRevLett.94.102501}.

The mass residual field is represented as a function $\Delta M(Z,N)$ defined on the proton--neutron plane. Prior to the Fourier analysis, the residual field is centered and normalized according to
\begin{equation}
\widetilde{\Delta M}(Z,N)
=
\frac{
\Delta M(Z,N)-\langle \Delta M\rangle
}{
\sigma_{\mathrm{rms}}
},
\end{equation}
where the root-mean-square fluctuation is defined as
\begin{equation}
\sigma_{\mathrm{rms}}
=
\sqrt{
\frac{1}{N_{\mathrm{nuc}}}
\sum_{Z,N}
\left[
\Delta M(Z,N)-\langle \Delta M\rangle
\right]^2 },
\end{equation}
with $N_{\mathrm{nuc}}$ denoting the total number of nuclei included in the analysis. The discrete two-dimensional Fourier transform of the normalized residual field is then given by
\begin{equation}
\widetilde{F}(k_Z,k_N)
=
\sum_{Z,N}
\widetilde{\Delta M}(Z,N)
\, e^{-2\pi i \left(
\frac{k_Z Z}{L_Z}
+
\frac{k_N N}{L_N}
\right)},
\end{equation}
where $L_Z$ and $L_N$ denote the lattice dimensions in the proton and neutron directions, respectively. The corresponding two-dimensional power spectrum is defined as
\begin{equation}
P(k_Z,k_N)
=
\frac{
|\widetilde{F}(k_Z,k_N)|^2
}{
N_{\mathrm{nuc}}
}.
\end{equation}
In this representation, low values of $(k_Z,k_N)$ correspond to slowly varying collective structures extending across broad regions of the nuclear chart, whereas large wave numbers probe localized fluctuations and rapidly varying correlations. The two-dimensional spectrum therefore provides direct information on the spatial organization of residual structures and their characteristic correlation scales in both proton and neutron directions simultaneously~\cite{Relano2002,PhysRevLett.93.244101,NIU2018759}.

Because experimentally known nuclei occupy an irregular and incomplete region of the $(Z,N)$ plane, constructing a regular lattice representation of the residual surface generally requires additional assumptions for unmeasured nuclei. In the present work, no artificial completion of the nuclear chart is introduced. Instead, the Fourier analysis is performed directly on the experimentally available nuclei through a discrete transform evaluated only at the measured $(Z,N)$ coordinates. This avoids distortions associated with auxiliary filling procedures. Assigning zero values to unmeasured lattice points would artificially enhance short-wavelength discontinuities and therefore generate spurious high-frequency spectral components, whereas filling missing regions through local averaging would artificially smooth the residual surface and introduce additional low-frequency coherence. By avoiding both procedures, the extracted spectra more faithfully reflect the intrinsic correlation structure contained in the available nuclear mass residuals.

To isolate directional dependencies in the residual correlations, projected power spectra are constructed by summing over one momentum direction,
\begin{equation}
P_Z(k_Z)
=
\sum_{k_N} P(k_Z,k_N),
\end{equation}
and
\begin{equation}
P_N(k_N)
=
\sum_{k_Z} P(k_Z,k_N).
\end{equation}
These projected spectra separately characterize correlations associated with variations along the proton and neutron directions. Enhanced low-frequency power in either projection signals the persistence of coherent structures preferentially aligned along the corresponding axis, while flatter spectra indicate fragmented or weakly correlated behavior~\cite{Relano2002,PhysRevLett.93.244101,NIU2018759}.

In addition to directional projections, an angle-averaged radial spectrum is introduced to characterize the isotropic distribution of spectral power. Defining the radial wave number as
\begin{equation}
k = \sqrt{k_Z^2 + k_N^2},
\end{equation}
the radial power spectrum is constructed by grouping all Fourier-grid points whose wave numbers lie within the same annular interval centered around $k$. Denoting by $\mathcal{A}_i$ the set of Fourier points belonging to the $i$-th annulus and by $N_i$ the total number of points inside that shell, the corresponding radial power is defined as
\begin{equation}
P_{\mathrm{radial}}(k_i)
=
\frac{1}{N_i}
\sum_{(k_Z,k_N)\in \mathcal{A}_i}
P(k_Z,k_N).
\end{equation}
This averaging procedure removes directional anisotropies while simultaneously normalizing the spectral contribution by the number of Fourier-grid points contained within each radial shell. Consequently, the resulting spectrum measures the mean spectral power associated with a characteristic spatial scale $k$, independent of the angular orientation in momentum space. The radial spectrum therefore isolates the global scale dependence of the residual fluctuations and provides a model-independent measure of the degree of long-range coherence remaining after NN filtering~\cite{Relano2002,PhysRevLett.93.244101,NIU2018759}.

\subsubsection*{Spectral rigidity and $\Delta_3$ statistics}

While Fourier analysis characterizes the distribution of correlations across different frequency scales, it does not directly quantify the degree of long-range rigidity present in the residual fluctuations. To probe the persistence of cumulative correlations over extended regions of the nuclear chart, a complementary analysis based on the Dyson--Mehta spectral rigidity statistic $\Delta_3$ is employed~\cite{Dyson1962III,Mehta2004}.

The ordered residual sequence $\Delta M(n)$ is first transformed into a cumulative fluctuation series defined as
\begin{equation}
X(m)
=
\sum_{n=1}^{m}
\Delta M(n),
\end{equation}
where the index $n$ follows the chosen boustrophedon ordering. In this representation, localized uncorrelated fluctuations tend to average out over long distances, whereas coherent residual correlations produce systematic deviations in the cumulative trajectory. The function $X(m)$ therefore provides a direct measure of the large-scale organization of the residual structure~\cite{Relano2002,PhysRevLett.93.244101}.

The spectral rigidity statistic $\Delta_3(L)$ is defined as the least-square deviation of the cumulative sequence from the best linear fit over an interval of length $L$,
\begin{equation}
\Delta_3(L)
=
\frac{1}{L}
\min_{A,B}
\int_{x}^{x+L}
\left[
X(m) - Am - B
\right]^2
dm,
\end{equation}
where the minimization with respect to the coefficients $A$ and $B$ removes the smooth linear trend within the interval. The resulting quantity measures the degree to which the cumulative fluctuations remain rigid or correlated over large scales~\cite{Dyson1962III,Mehta2004,Guhr1998}.

For completely uncorrelated fluctuations, $\Delta_3(L)$ increases rapidly with interval length, reflecting the absence of long-range coherence. In contrast, correlated or rigid spectra exhibit slower growth, indicating the persistence of structured collective behavior across extended regions of the sequence. Intermediate behavior corresponds to partially correlated systems in which coherent structures persist only over restricted scales~\cite{Mehta2004,Guhr1998,RevModPhys.81.539}.

In the present framework, the $\Delta_3$ statistic is evaluated for both the original mass residuals and the residuals remaining after NN filtering. Comparing the evolution of $\Delta_3(L)$ before and after filtering therefore provides a complementary diagnostic of how efficiently long-range correlations are removed by the hybrid model. A substantial suppression of spectral rigidity indicates successful elimination of coherent systematic trends, whereas residual deviations from random-matrix-like behavior signal the persistence of nontrivial many-body correlations not fully captured within the adopted theoretical framework~\cite{PhysRevLett.96.042502,MOLINARI200648,Aberg2002}.

\subsection{Hierarchical residual decomposition}
The starting point of the multistage analysis is the mass residual $\Delta M$, defined as the difference between experimental binding energies and the predictions of a reference mass model. These residuals encode both smooth systematic deviations and localized irregularities arising from missing or incomplete physics. NNs are used as active spectral filters designed to systematically suppress these correlated trends from nuclear mass residuals, allowing us to study the capacity threshold required to drive the remaining fluctuations to the white-noise limit.

The networks are trained as nonlinear regression models using the proton and neutron numbers as inputs~\cite{Goodfellow2016,Bishop2006}. Under strongly regularized training conditions, they predominantly capture the smooth, long-range correlations of the residual surface. These structures typically correspond to broad collective trends, deformation effects, and other systematic contributions that extend across neighboring nuclei. By progressively relaxing the regularization, we increase the model capacity, allowing the network to absorb and suppress progressively shorter-range chaotic fluctuations that otherwise resist low-capacity modeling.

\subsubsection*{Hierarchical NN filtering of mass residuals}

To remove correlated structures from the nuclear mass residuals, the HRD framework trains fully connected FFNNs and MoE architectures successively on the residual field. At each stage, the NN targets only the correlations remaining after the previous stage.

Starting from the initial residual field $\Delta M$, the first NN produces a nonlinear approximation $f_1(Z,N)$. The remaining residual is then
\begin{equation}
\Delta M_1
=
\Delta M - f_1(Z,N).
\end{equation}

The same procedure is repeated recursively according to
\begin{equation}
\Delta M_i
=
\Delta M_{i-1} - f_i(Z,N),
\end{equation}
where each NN is trained on the residual field obtained from the previous stage. The early stages mainly remove the smoother and more strongly correlated parts of the residual surface. The later stages instead act on weaker and more localized fluctuations. In practice, the hierarchy contains a large number of NNs trained one after another in series. Depending on the stopping conditions and the regularization schedule, the total number of stages can range from $\sim100$ to several hundred NNs in a single training sequence~\cite{Zai2026CoNN,PhysRevC.111.034305}.

The training starts with relatively strong $L_2$ regularization and gradually moves toward weaker regularization strengths. Typical values decrease from about $L_2\sim10^{-1}$ to $L_2\sim10^{-7}$. Under strong regularization, the NNs are restricted to learning broad smooth correlations extending over large regions of the nuclear chart. As the regularization is relaxed to extremely small values, the model capacity increases, allowing the networks to resolve highly localized and short-range variations. The scale of regularization required to completely flatten the spectrum acts as a physical probe, indicating the transition point where structured many-body correlations are fully resolved and only uncorrelated noise remains.

Each stage is trained using an adaptive plateau stopping criterion. A maximum epoch limit is imposed, but the actual stopping point depends on how the training loss evolves. If the variation in the loss remains below a fixed threshold for several consecutive epochs, the training stage is stopped and the next residual-learning step is started. Since the residual structure becomes weaker in the later stages, this criterion helps avoid unnecessary over-training and also improves numerical stability.

Additional stabilization is introduced through weighted residual training. Residual points with exceptionally large magnitudes are given smaller weights during optimization so that isolated outliers do not dominate the fitting process. The NNs therefore remain mainly focused on correlated structures shared among neighboring nuclei rather than on strongly localized fluctuations.

The FFNN stages provide a global nonlinear approximation to the residual field. Each stage consists of a deep NN with several hidden layers and nonlinear activation functions~\cite{Goodfellow2016,Bishop2006}. Since all nuclei share a single global mapping, the FFNN hierarchy primarily removes broad smooth structures extending over large regions of the nuclear chart.

The residual structure, however, is highly heterogeneous and can differ strongly between shell regions, deformation regions, and transitional nuclei. Because of this, MoE architectures~\cite{Jacobs1991,Jordan1994} are also included in the residual-learning hierarchy. In the MoE framework, several subnetworks contribute together to the residual prediction through an adaptive gating function. The resulting prediction is written as
\begin{equation}
f_{\mathrm{MoE}}(Z,N)
=
\sum_{a=1}^{K}
g_a(Z,N)\, f_a(Z,N),
\end{equation}
where $f_a(Z,N)$ denotes the prediction of the $a$-th subnetwork and the gating weights satisfy
\begin{equation}
\sum_{a=1}^{K} g_a(Z,N)=1.
\end{equation}

Different subnetworks then become specialized for different regions of the residual field. In practice, this includes shell-dominated nuclei, deformation regions, pairing-sensitive sectors, and more localized fluctuation patterns. Entropy regularization is also applied to the gating probabilities to discourage mode collapse toward a single effective global mapping and maintain expert diversity.

Both FFNN and MoE hierarchies are trained independently for each reference mass model. The resulting residual sequences are later analyzed using Fourier spectra and spectral-rigidity statistics in order to study how the residual correlations evolve during the successive filtering stages.

\subsubsection*{Training and testing protocol}

The NN residual-learning framework is applied to experimental nuclear mass datasets containing approximately $2000$--$2500$ nuclei, depending on the reference mass model and the adopted selection criteria. Because the statistical properties and dominant correlation structures vary substantially between light and heavy nuclei, the dataset is initially partitioned into two broad mass regions defined by
$$A \leq 120,
\qquad
A > 120.
$$This division allows the NNs to learn correlation patterns within more homogeneous sectors of the nuclear chart and reduces the mixing of structurally distinct physical regimes during training.

Within each mass region, the nuclei are randomly partitioned into $22$ independent groups of approximately equal size. For a given major training run, one group is reserved exclusively for testing, while the remaining $21$ groups are used for training the NN hierarchy. The complete residual-learning framework, including all sequential FFNN and MoE stages, is then trained only on the nuclei belonging to the training subset.

After completion of the training sequence, predictions are generated for the nuclei contained in the excluded test group. The procedure is subsequently repeated $22$ times, each time selecting a different group as the testing subset. In this way, every nucleus in the dataset is predicted exactly once as a held-out test point.

This protocol ensures that the predicted residual of a given nucleus is obtained without allowing the NNs to directly access the corresponding experimental residual during training. Only neighboring nuclei and other correlated structures present in the remaining training groups are available to the learning procedure. Consequently, the final predictions correspond to out-of-sample residual estimates rather than interpolations of previously observed targets~\cite{PhysRevResearch.2.043363,Huang2025}.

The complete collection of predictions from all $22$ testing cycles is subsequently recombined to reconstruct the full predicted residual field over the nuclear chart. This reconstructed dataset is then used for the Fourier and spectral-rigidity analyses described in the following sections.

\subsection{PINE Model}\label{sec:pine}

To reduce model-dependent biases and improve the robustness of the residual predictions, the HRD framework is applied independently to three different theoretical nuclear mass models: the Duflo--Zuker (DZ10), finite-range droplet model (FRDM), and Hartree--Fock--Bogoliubov (HFB24) mass models~\cite{DZ10,qi2015theoretical,FRDM2012,HFB24}. For each of these three models, both the FFNN and MoE residual-learning architectures are trained separately following the multistage decomposition procedure described above.

This construction produces six independent NN-corrected residual models,
\begin{align*}
\Delta M_{\mathrm{DZ}}^{\mathrm{FFNN}},
\qquad
\Delta M_{\mathrm{DZ}}^{\mathrm{MoE}},
\qquad
\Delta M_{\mathrm{FRDM}}^{\mathrm{FFNN}},
\\[0.5ex]
\Delta M_{\mathrm{FRDM}}^{\mathrm{MoE}},
\qquad
\Delta M_{\mathrm{HFB}}^{\mathrm{FFNN}},
\qquad
\Delta M_{\mathrm{HFB}}^{\mathrm{MoE}}.
\end{align*}
Each of these corrected residual fields captures partially overlapping but nonidentical correlation structures. Since different theoretical mass models contain different systematic deficiencies and encode nuclear structure effects in different ways, the residual correlations learned by the NNs are not identical across the six constructions. Combining the resulting predictions therefore allows complementary information from distinct physical descriptions and NN architectures to be incorporated simultaneously.

The final hybrid residual predictor developed in the present work is constructed through a weighted neural ensemble of these six independently filtered residual fields. We refer to this final framework as the PINE model. The PINE construction combines complementary information from multiple global nuclear mass models and neural residual-learning architectures to progressively suppress model-specific systematic correlations while preserving residual structures that are robust across independent descriptions.

The final PINE residual prediction is constructed as a weighted linear combination of the six corrected residual models,
\begin{align*}
\Delta M_{\mathrm{PINE}}
={}&
w_1 \Delta M_{\mathrm{DZ}}^{\mathrm{FFNN}}
+
w_2 \Delta M_{\mathrm{DZ}}^{\mathrm{MoE}}
\\[0.5ex]
&+
w_3 \Delta M_{\mathrm{FRDM}}^{\mathrm{FFNN}}
+
w_4 \Delta M_{\mathrm{FRDM}}^{\mathrm{MoE}}
\\[0.5ex]
&+
w_5 \Delta M_{\mathrm{HFB}}^{\mathrm{FFNN}}
+
w_6 \Delta M_{\mathrm{HFB}}^{\mathrm{MoE}}.
\end{align*}
The coefficients are given by
\begin{align*}
w_1 &= 0.0957,
&
w_2 &= 0.2414,
&
w_3 &= 0.1508,
\\
w_4 &= 0.1865,
&
w_5 &= 0.1726,
&
w_6 &= 0.1529.
\end{align*}
The coefficients are determined by minimizing the overall prediction error of the combined model subject to normalization constraints. Since the six NN constructions exhibit distinct residual correlation patterns and systematic behaviors across the nuclear chart, the weighted ensemble suppresses model-specific fluctuations while retaining structures that are consistently supported across multiple independent descriptions. Residual refinement and model-repair strategies have also been successfully employed to improve the predictive performance of global nuclear mass models~\cite{PhysRevC.111.024306,PhysRevC.111.054322}.

The resulting PINE model therefore combines the global smoothing capability of FFNN residual learning, the localized adaptive structure identification of MoE networks, and the complementary physical information contained within the DZ, FRDM, and HFB mass models. In this sense, the PINE framework represents the final stage of the hierarchical residual decomposition developed in the present work, integrating multiple theoretical descriptions and learning strategies into a unified residual predictor. The residual field obtained from the PINE construction is subsequently analyzed through Fourier spectra and spectral-rigidity diagnostics in order to characterize the correlation structures that survive after the complete hybrid filtering procedure.

\section{Results}
\subsection{One-dimensional spectral analysis of the individual hybrid residual models}

To investigate the evolution of residual correlations after NN filtering, the reconstructed hybrid residual sequences are analyzed using one-dimensional Fourier power spectra constructed from two different boustrophedon orderings: a mass-based ordering and a shell-based ordering~\cite{PhysRevLett.94.102501,Hirsch2004-sl,Hirsch2005-rr}. The former corresponds to the conventional mass-based boustrophedon sequence employed in previous spectral studies of nuclear mass fluctuations~\cite{PhysRevLett.94.102501,Hirsch2004-sl}, whereas the latter reorganizes the nuclei according to the shell structure.
Both orderings are introduced to probe complementary aspects of the residual dynamics. The mass-based ordering primarily emphasizes large-scale global trends distributed across broad regions of the nuclear chart. In contrast, the shell-based ordering reorganizes the nuclei according to their proximity to major shell closures and therefore isolates correlations associated with shell structure and shell-distance hierarchies~\cite{BohrMottelsonI,BohrMottelsonII}.

The NN residual-learning framework is applied independently to the DZ, FRDM, and HFB mass models using both FFNN and MoE architectures, and the resulting spectral evolution is examined below. The corresponding prediction errors before and after NN filtering are summarized in Table~\ref{tab:rmse_individual_models}. The Fourier spectral slopes extracted from the mass-based and shell-based boustrophedon orderings are summarized in Tables~\ref{tab:slope_mass_based} and~\ref{tab:slope_shell_based}, respectively.
%==============================================================
% RMSE TABLE
%==============================================================

\begin{table}[H]
\caption{Root-mean-square deviations before and after NN residual learning for the six hybrid constructions.}
\label{tab:rmse_individual_models}
\begin{ruledtabular}
\begin{tabular}{lcc}
Model & Before (MeV) & After (MeV) \\
\hline
DZ--FFNN   & 0.598 & 0.253 \\
DZ--MoE    & 0.598 & 0.251 \\
FRDM--FFNN & 0.596 & 0.249 \\
FRDM--MoE  & 0.596 & 0.249 \\
HFB--FFNN  & 0.541 & 0.318 \\
HFB--MoE   & 0.541 & 0.329 \\
\end{tabular}
\end{ruledtabular}
\end{table}

\begin{table}[H]
\caption{Spectral slopes obtained from the mass-based boustrophedon ordering before and after NN residual learning.}
\label{tab:slope_mass_based}
\begin{ruledtabular}
\begin{tabular}{lcc}
Model & Before & After \\
\hline
DZ--FFNN   & $-0.954$ & $-0.136$ \\
DZ--MoE    & $-0.954$ & $-0.061$ \\
FRDM--FFNN & $-0.848$ & $0.197$ \\
FRDM--MoE  & $-0.848$ & $0.282$ \\
HFB--FFNN  & $-0.556$ & $-0.051$ \\
HFB--MoE   & $-0.556$ & $0.006$ \\
\end{tabular}
\end{ruledtabular}
\end{table}

%==============================================================
% SHELL-BASED SLOPE TABLE
%==============================================================

\begin{table}[H]
\caption{Spectral slopes obtained from the shell-based boustrophedon ordering before and after NN residual learning.}
\label{tab:slope_shell_based}
\begin{ruledtabular}
\begin{tabular}{lcc}
Model & Before & After \\
\hline
DZ--FFNN   & $-0.467$ & $-0.002$ \\
DZ--MoE    & $-0.467$ & $-0.009$ \\
FRDM--FFNN & $-0.592$ & $-0.178$ \\
FRDM--MoE  & $-0.592$ & $-0.101$ \\
HFB--FFNN  & $-0.284$ & $-0.001$ \\
HFB--MoE   & $-0.284$ & $0.027$ \\
\end{tabular}
\end{ruledtabular}
\end{table}

Prior to NN filtering, the DZ and FRDM residuals exhibit large negative spectral slopes in both orderings, indicating that strong low-frequency structures remain present in the residual field. These structures primarily originate from slowly varying systematic effects extending over large regions of the nuclear chart.

\begin{figure}[H]
\centering
\includegraphics[width=0.98\columnwidth]{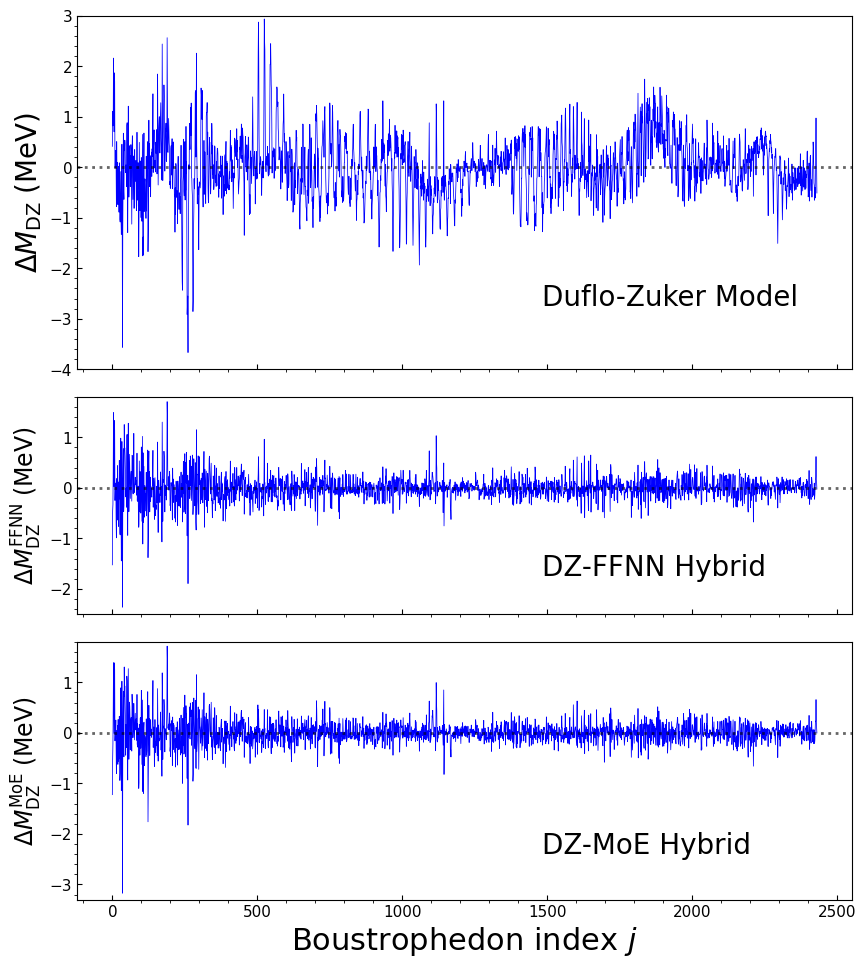}
\caption{Residual sequences for the DZ model before and after FFNN and MoE residual learning using the {mass-based boustrophedon ordering}.}
\label{fig:dz_fig1}
\end{figure}

\begin{figure}[H]
\centering
\includegraphics[width=0.98\columnwidth]{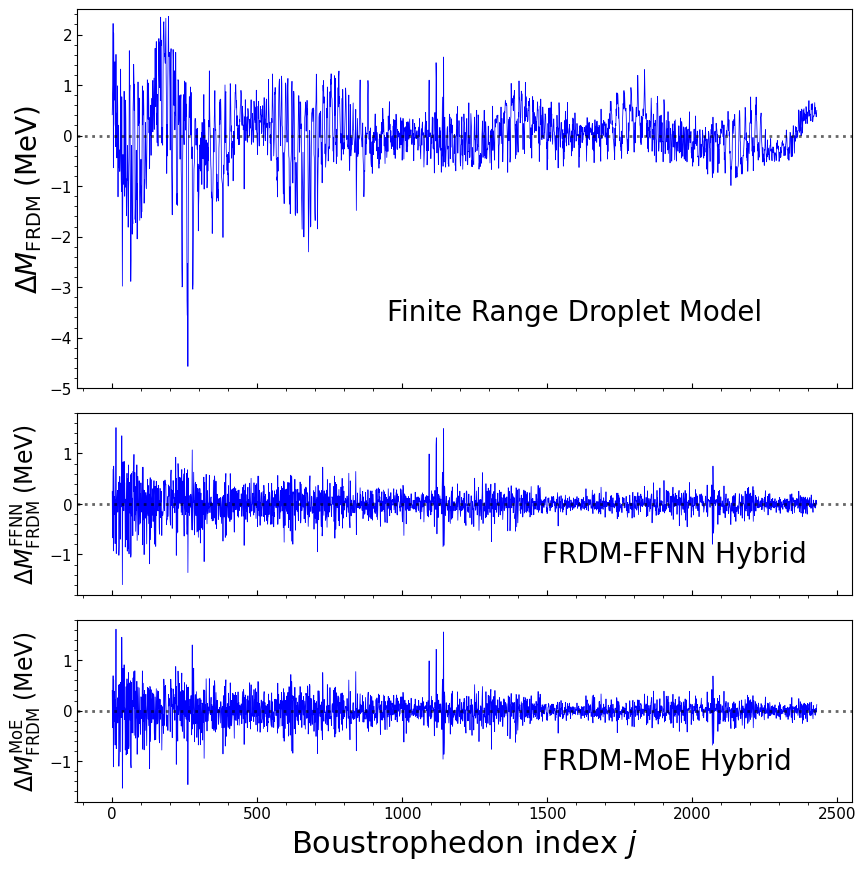}
\caption{Residual sequences for the FRDM model before and after FFNN and MoE residual learning using the {mass-based boustrophedon ordering}.}
\label{fig:frdm_fig1}
\end{figure}

\begin{figure}[H]
\centering
\includegraphics[width=0.98\columnwidth]{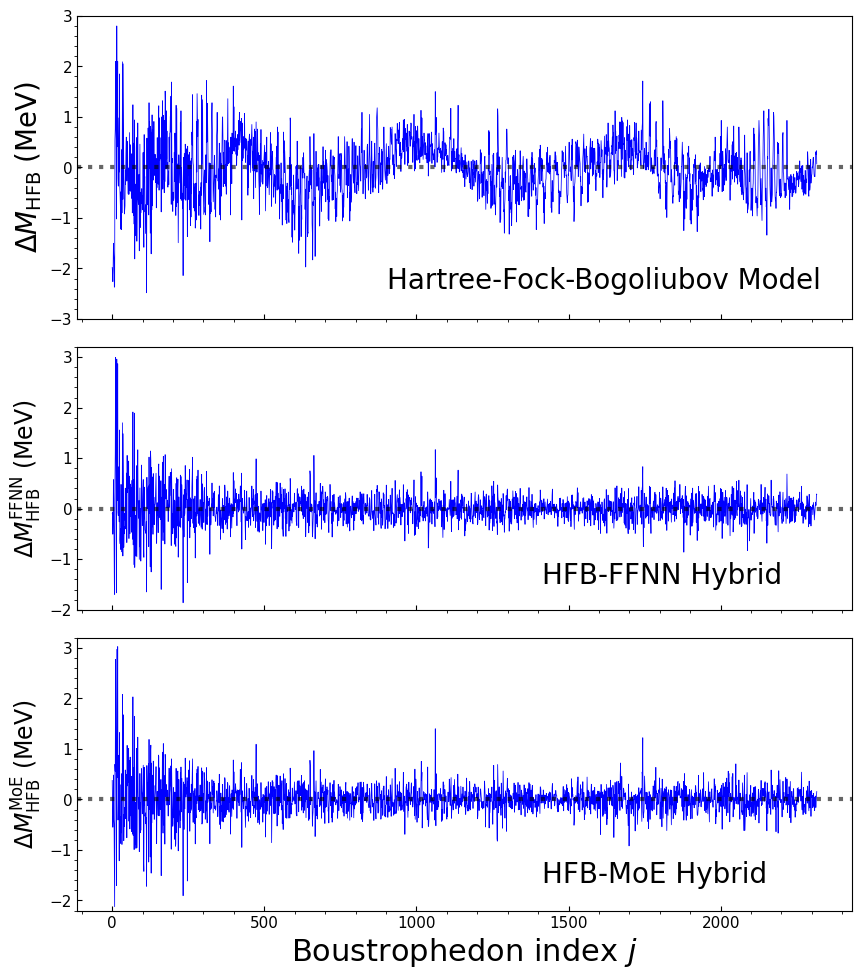}
\caption{Residual sequences for the HFB model before and after FFNN and MoE residual learning using the {mass-based boustrophedon ordering}.}
\label{fig:hfb_fig1}
\end{figure}

After the FFNN and MoE corrections are introduced, the spectral slopes collapse toward values much closer to zero. Simultaneously, the Fourier spectra become considerably flatter across the full frequency interval. This transition demonstrates that the HRD hierarchy, acting as a high-pass spectral filter, efficiently absorbs and suppresses the chaotic $1/f$ correlations from the residuals, leaving behind flat, uncorrelated white-noise fluctuations.

The shell-based ordering exhibits systematically smaller initial slopes compared with the mass-based ordering. This indicates that a substantial fraction of the strongest long-range correlations originates from broad global mass trends extending across the nuclear chart rather than from purely shell-localized effects~\cite{PhysRevLett.94.102501,Hirsch2004-sl,NIU2018759}. 

Most importantly, after NN residual learning the dominant low-frequency structures and quantum-chaotic rigidity signatures are almost completely suppressed in both the DZ and HFB constructions. Their post-filtered spectra become nearly flat in both the mass-based and shell-based orderings. This demonstrates that a high-capacity neural filter has the explicit capacity to systematically eliminate $1/f$ correlations and flatten the spectrum to the white-noise limit~\cite{Relano2002,PhysRevLett.93.244101,PhysRevLett.118.204101}.

\begin{figure}[H]
\centering
\includegraphics[width=0.98\columnwidth]{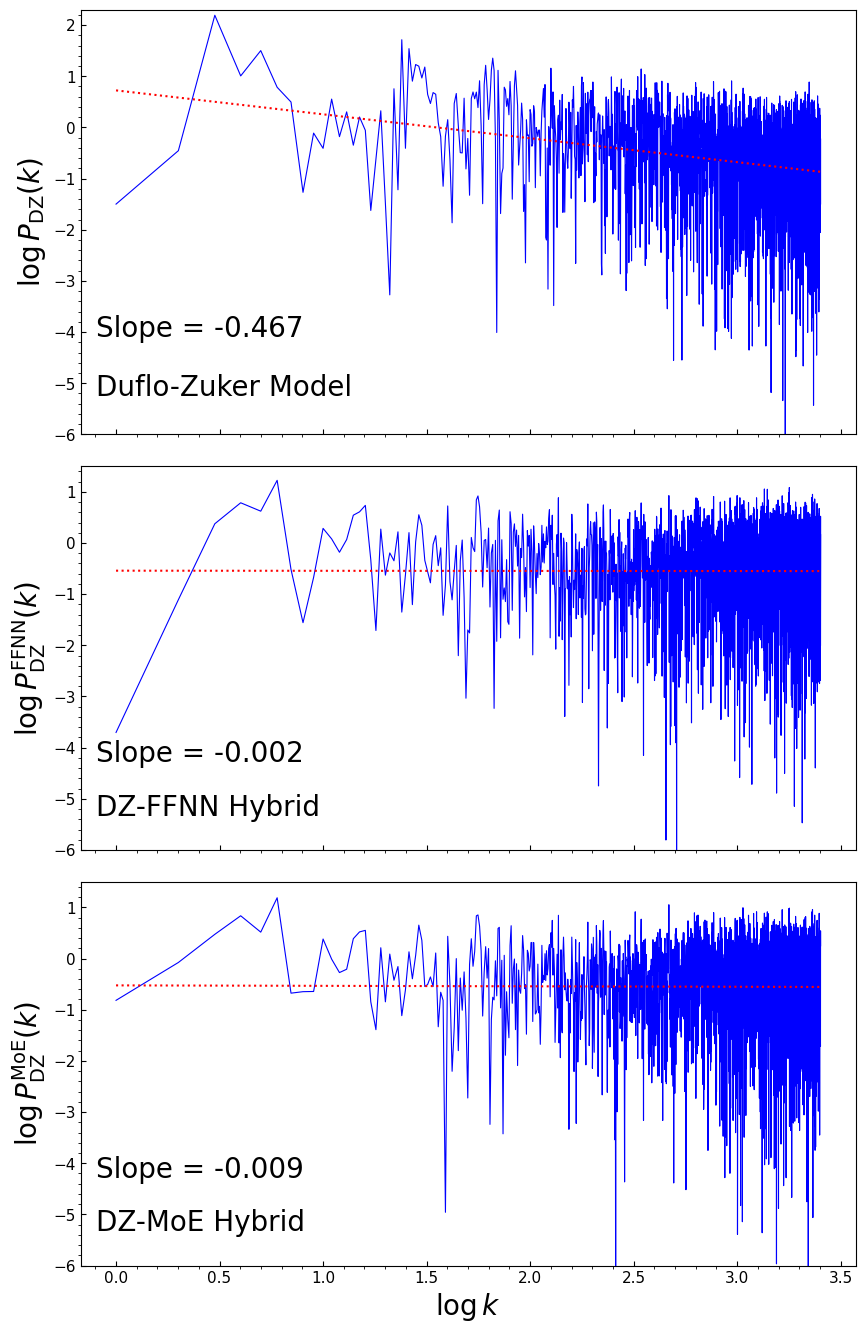}
\caption{Fourier power spectra corresponding to the {shell-based boustrophedon ordering} for the DZ residual sequences before and after FFNN and MoE residual learning.}
\label{fig:dz_fig2_shell}
\end{figure}

The FRDM residuals show somewhat different behavior. Even after filtering, a weak low-frequency structure still remains. The higher-frequency region also becomes more visible after filtering, especially in the mass-based ordering. Most of the remaining fluctuations are now more local and fragmented after the smooth correlated structures are removed~\cite{FRDM2012,Hirsch2005-rr,MOLINARI200648}.

The shell-based spectra also become much flatter after the filtering stages. Once the dominant shell-related structures are removed, the residual fluctuations away from shell closures show only weak large-scale correlations~\cite{BohrMottelsonI,BohrMottelsonII,PhysRevLett.94.102501}. Most of the remaining residuals are now dominated by local fluctuations and weaker many-body effects that are not fully removed within the present hybrid description~\cite{Aberg2002,PhysRevLett.96.042502,MOLINARI200648}.

\begin{figure}[H]
\centering
\includegraphics[width=0.98\columnwidth]{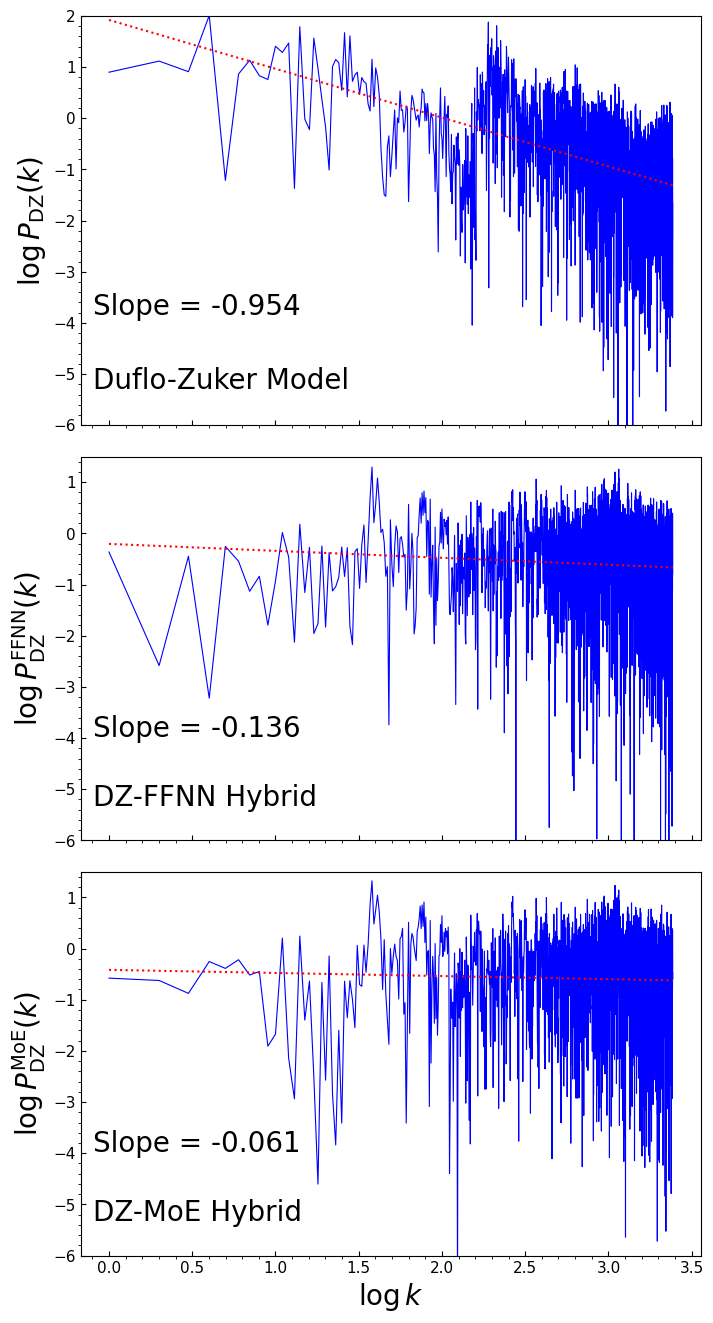}
\caption{Fourier power spectra corresponding to the {mass-based boustrophedon ordering} for the DZ residual sequences before and after FFNN and MoE residual learning.}
\label{fig:dz_fig2}
\end{figure}

The DZ residuals show strong correlated structures extending across large regions of the boustrophedon sequence, as seen in Fig.~\ref{fig:dz_fig1}. The Fourier spectra in Figs.~\ref{fig:dz_fig2} and~\ref{fig:dz_fig2_shell} are dominated by low-frequency components and large negative slopes. After the NN filtering stages, the residual amplitudes become much smaller and the spectra become flatter. Only weak low-frequency structures still remain.

\begin{figure}[H]
\centering
\includegraphics[width=0.98\columnwidth]{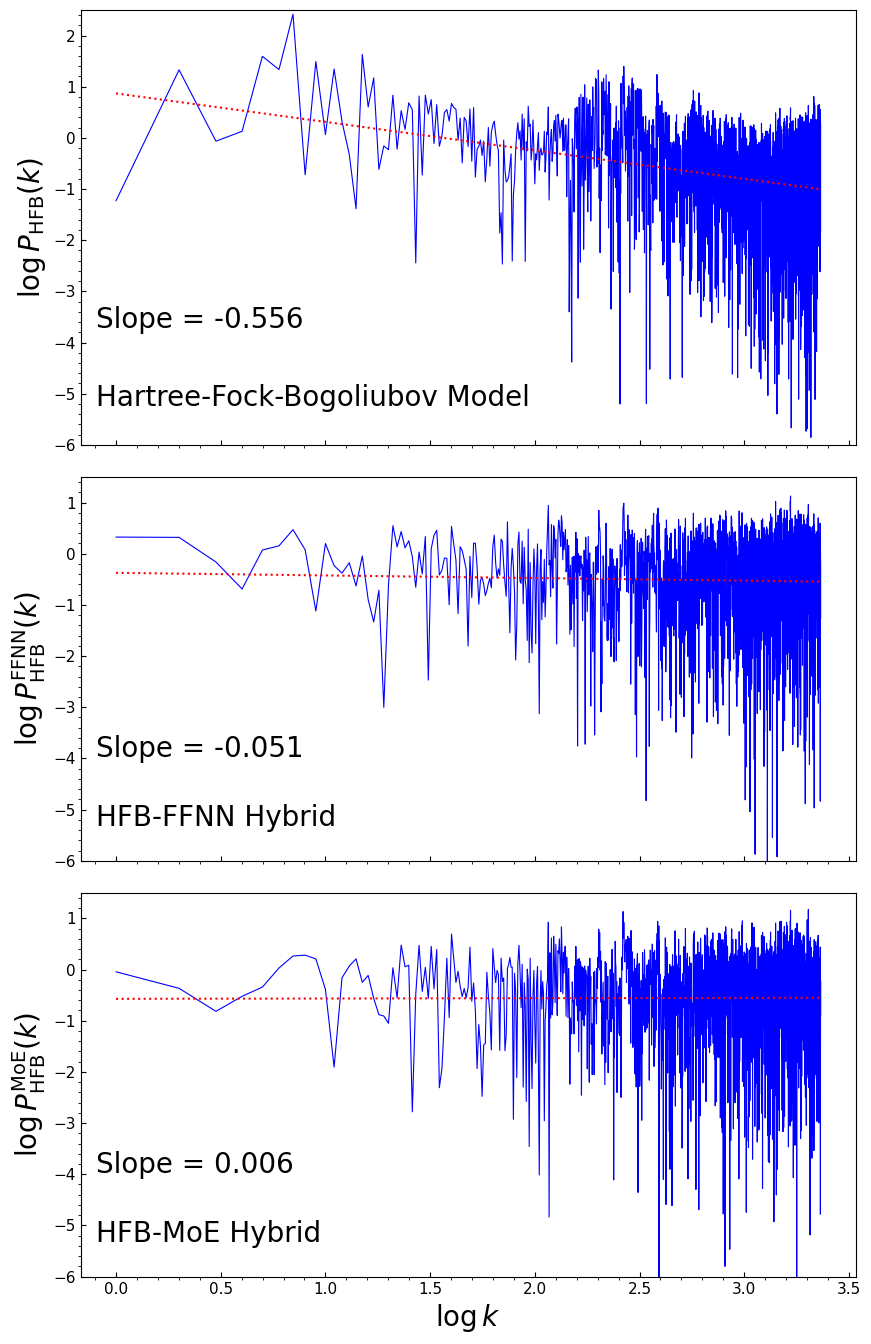}
\caption{Fourier power spectra corresponding to the {mass-based boustrophedon ordering} for the HFB residual sequences before and after FFNN and MoE residual learning.}
\label{fig:hfb_fig2}
\end{figure}

\begin{figure}[H]
\centering
\includegraphics[width=0.98\columnwidth]{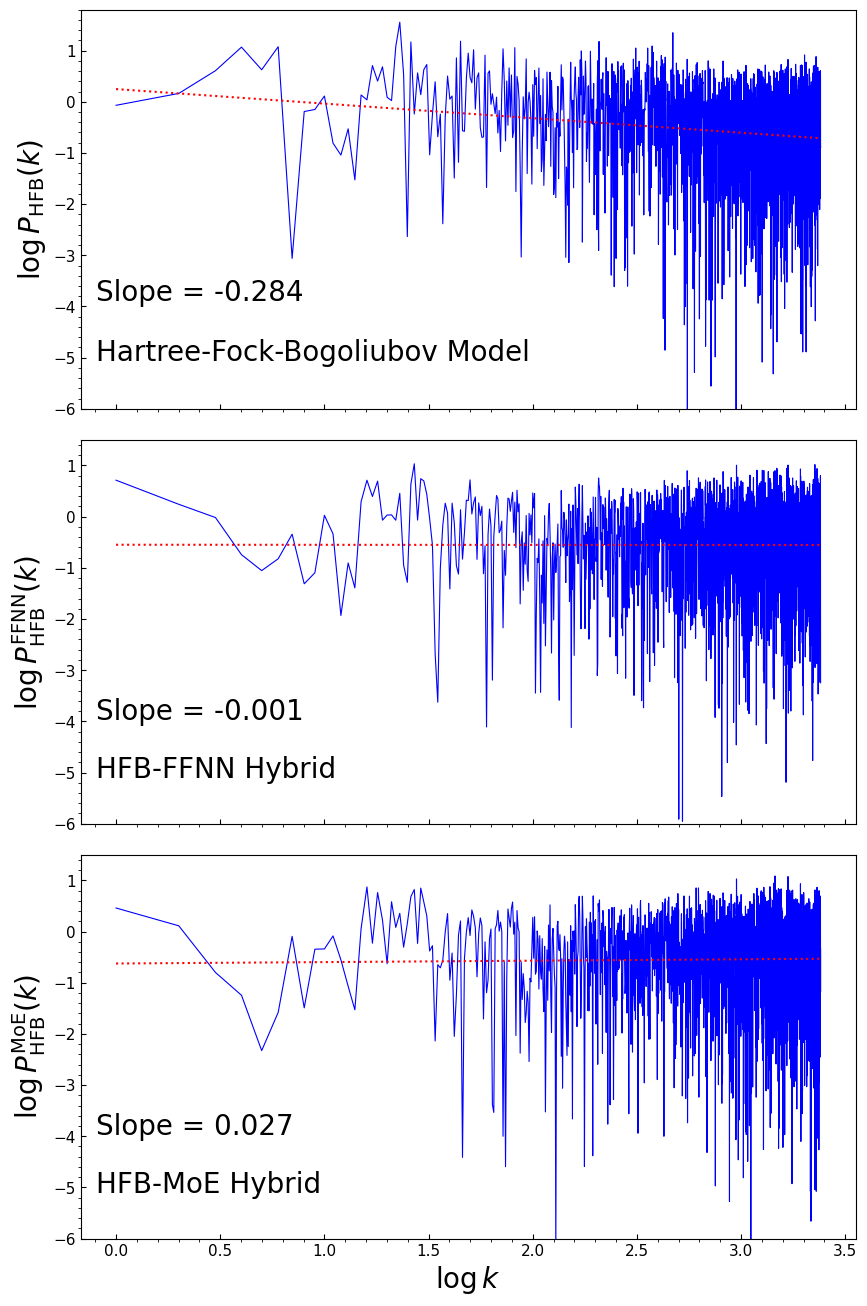}
\caption{Fourier power spectra corresponding to the {shell-based boustrophedon ordering} for the HFB residual sequences before and after FFNN and MoE residual learning.}
\label{fig:hfb_fig2_shell}
\end{figure}

\begin{figure}[H]
\centering
\includegraphics[width=0.98\columnwidth]{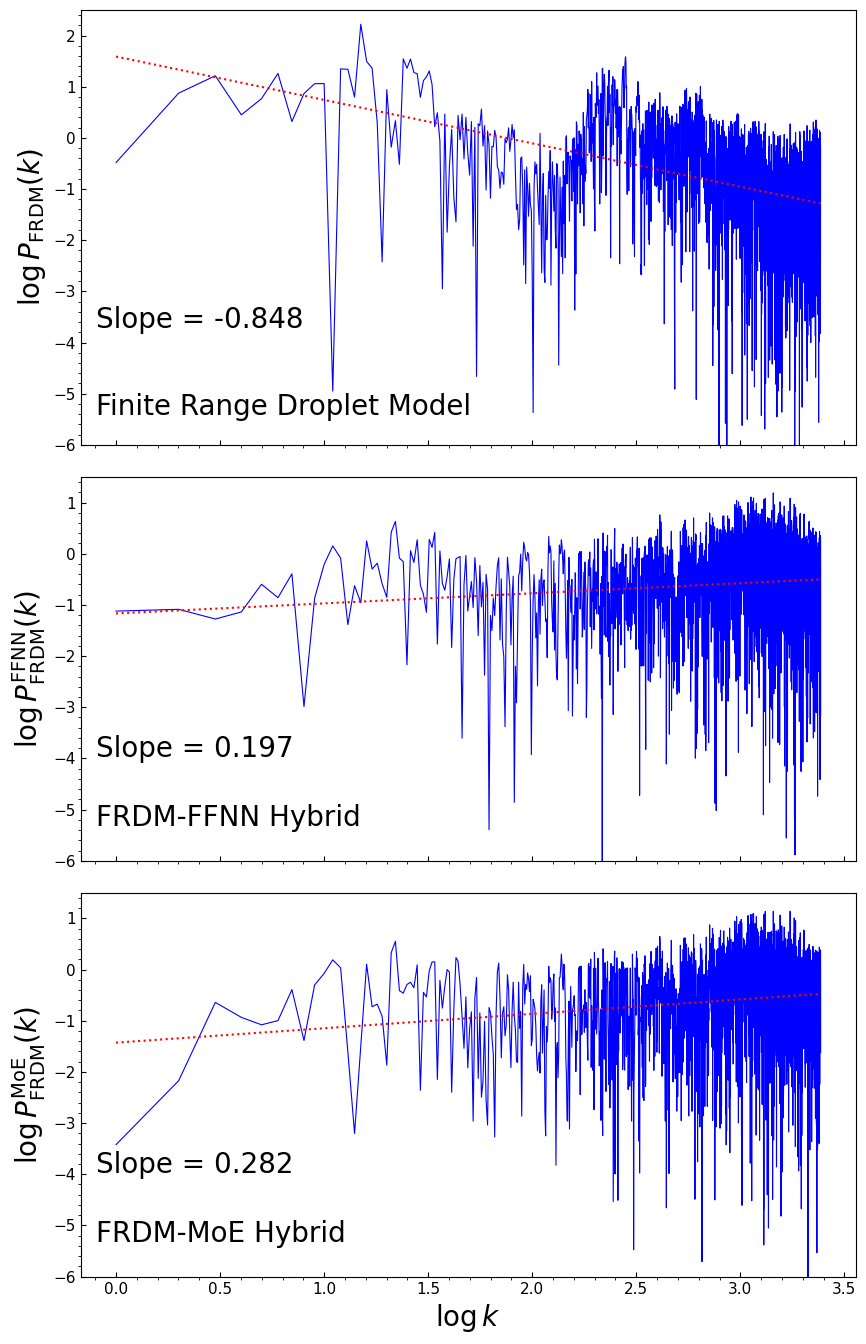}
\caption{Fourier power spectra corresponding to the {mass-based boustrophedon ordering} for the FRDM residual sequences before and after FFNN and MoE residual learning.}
\label{fig:frdm_fig2}
\end{figure}

\begin{figure}[H]
\centering
\includegraphics[width=0.98\columnwidth]{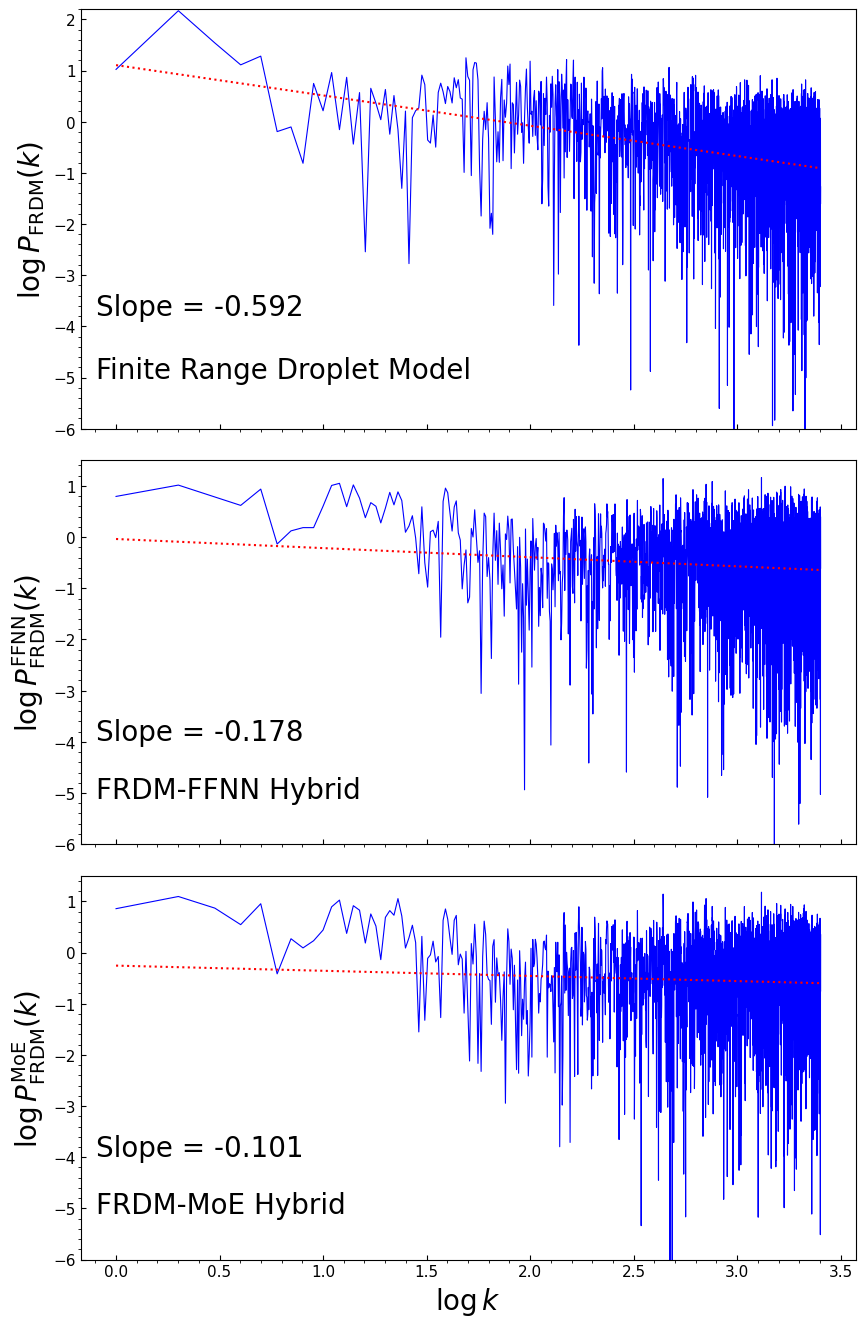}
\caption{Fourier power spectra corresponding to the {shell-based boustrophedon ordering} for the FRDM residual sequences before and after FFNN and MoE residual learning.}
\label{fig:frdm_fig2_shell}
\end{figure}

Among the three theoretical descriptions, the FRDM residuals exhibit comparatively strong low-frequency correlations prior to NN filtering, as shown in the residual sequences of Fig.~\ref{fig:frdm_fig1} and reflected in the pronounced negative slopes of the Fourier spectra in Figs.~\ref{fig:frdm_fig2} and~\ref{fig:frdm_fig2_shell}. After the FFNN and MoE residual-learning stages, the spectra flatten substantially. In the mass-based boustrophedon ordering, the post-filtered spectra additionally develop enhanced intermediate- and high-frequency power, leading to weakly positive spectral slopes and indicating that the remaining fluctuations are predominantly localized. In the shell-based ordering, however, weak negative slopes persist after filtering, suggesting that a fraction of the residual correlations remains associated with shell-distance-dependent structures.

The HFB residuals already show weaker long-range structures before NN filtering, as seen in Fig.~\ref{fig:hfb_fig1}. This can also be seen from the smaller initial spectral slopes in Figs.~\ref{fig:hfb_fig2} and~\ref{fig:hfb_fig2_shell} compared with the DZ and FRDM cases. Because of this, the reduction of the RMSE values after filtering is more moderate. Nevertheless, both the FFNN and MoE stages remove most of the remaining low-frequency power, and the final spectra become much flatter and closer to a weakly correlated regime.

The Fourier spectra after residual learning show that the dominant changes occur at low frequencies. The NN hierarchy removes most of the smooth structures extending across broad regions of the nuclear chart. These long-range structures are associated with systematic trends related to mean-field effects, shell evolution, and collective contributions in the underlying mass models~\cite{Bender2003,Caurier2005,Lunney2003}.

At higher frequencies, however, the spectra remain comparatively fragmented and irregular even after filtering. This behavior indicates that the residual-learning procedure does not simply suppress the overall residual amplitude uniformly across all scales, but instead selectively removes the coherent long-wavelength structures while leaving behind localized fluctuations with considerably weaker correlations. The surviving high-frequency components are therefore consistent with residual local irregularities arising from unresolved many-body effects, shell-fragmentation phenomena, deformation competition, and stochastic contributions that do not organize coherently over large regions of the nuclear chart~\cite{Aberg2002,MOLINARI200648,PhysRevLett.96.042502}.

\subsection{Spectral Analysis of the PINE Model}

Having introduced the PINE framework in Sec.~\ref{sec:pine}, we now investigate the statistical structure of the resulting residual fluctuations through one-dimensional and two-dimensional spectral analyses. The purpose of this section is to determine whether the remaining mass deviations in the PINE model preserve coherent low-frequency structures or whether the fluctuations approach an uncorrelated noise-like regime after the neural residual corrections.

Before proceeding to the spectral characterization, we summarize the overall predictive improvement obtained through the final residual-learning framework. The initial residual field extracted directly from the experimental data relative to the averaged theoretical construction, defined as the simple arithmetic average of the DZ10, FRDM, and HFB24 mass models prior to neural residual learning, exhibits a global root-mean-square deviation $\sigma_{\mathrm{rms}}\approx0.379~\mathrm{MeV}$. After applying the complete HRD residual-learning and weighted PINE hybridization procedure, the remaining residual fluctuations are reduced to $\sigma_{\mathrm{rms}}\approx0.184~\mathrm{MeV}$. This level of accuracy approaches the sub-$200~\mathrm{keV}$ regime discussed in high-precision hybrid nuclear mass studies~\cite{NIU2018759,universe7050131}.

\subsubsection*{One-dimensional Fourier analysis using mass-based boustrophedon ordering}

\begin{figure}[H]
\centering
\includegraphics[width=\columnwidth]{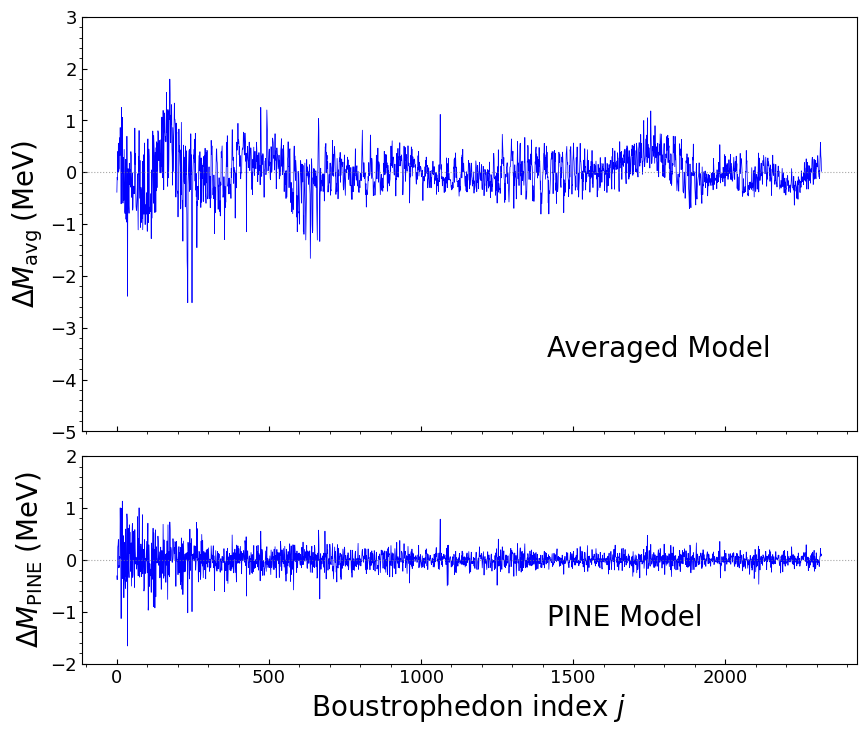}
\caption{
Residual fluctuations corresponding to the mass-based boustrophedon ordering. The upper panel corresponds to the averaged model, while the lower panel corresponds to the PINE-enhanced hybrid model.
}
\label{fig:pine_bous_residuals}
\end{figure}

\begin{figure}[H]
\centering
\includegraphics[width=\columnwidth]{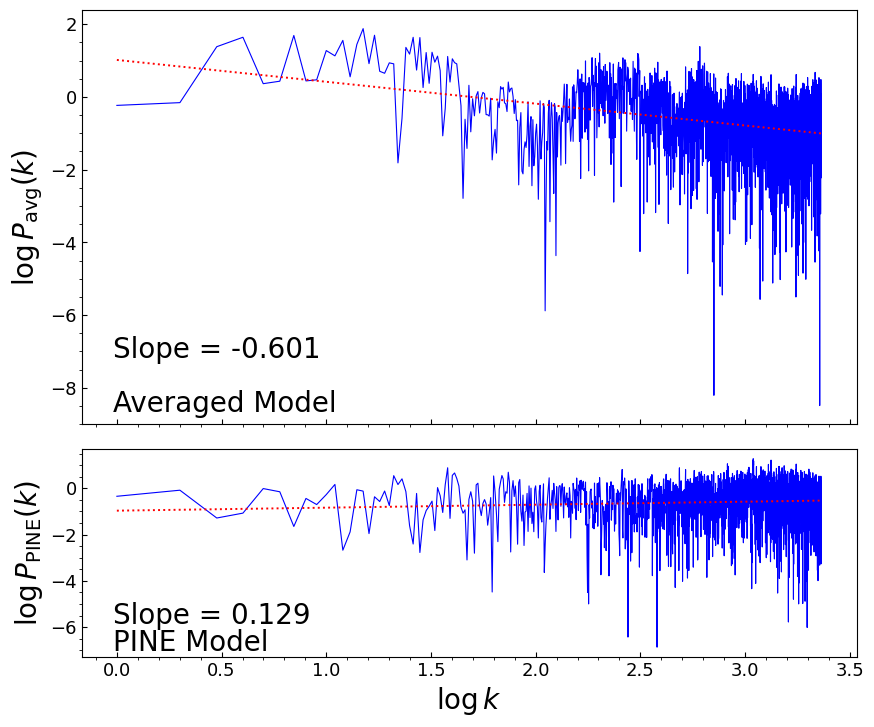}
\caption{
Fourier power spectra corresponding to the mass-based boustrophedon ordering.
}
\label{fig:pine_bous_fft}
\end{figure}

The first analysis is performed using the $A$-based boustrophedon ordering of nuclei, with the residual sequences shown in Fig.~\ref{fig:pine_bous_residuals}. Here, the ``averaged model'' refers to the simple arithmetic average of the three underlying theoretical mass models, DZ10, FRDM, and HFB24, prior to the application of neural residual learning.
The corresponding Fourier power spectra are displayed in Fig.~\ref{fig:pine_bous_fft}. The averaged model exhibits a pronounced low-frequency dominance with spectral slope $\approx -0.601$, indicating the presence of coherent slowly varying global structures across the nuclear chart. After application of the PINE correction, the Fourier spectrum becomes considerably flatter, with spectral slope $\approx 0.129$, demonstrating that most coherent long-range correlations have been suppressed by the neural residual learning.
\FloatBarrier

\subsubsection*{One-dimensional Fourier analysis using shell-based boustrophedon ordering}

\begin{figure}[H]
\centering
\includegraphics[width=0.98\columnwidth]{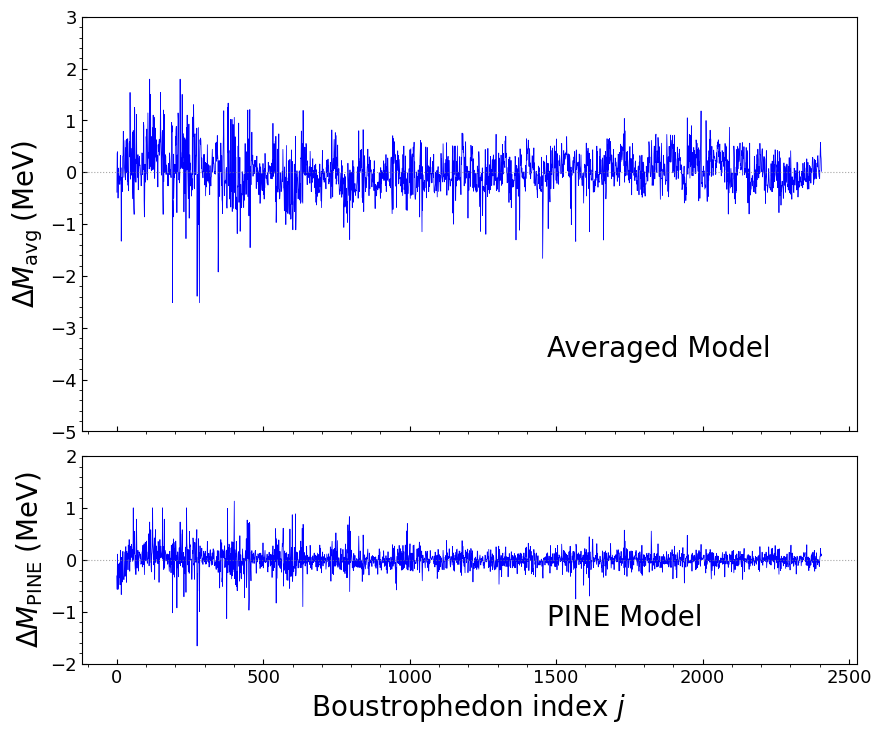}
\caption{
Residual fluctuations corresponding to the shell-based boustrophedon ordering.
}
\label{fig:pine_shell_residuals}
\end{figure}

\begin{figure}[H]
\centering
\includegraphics[width=0.98\columnwidth]{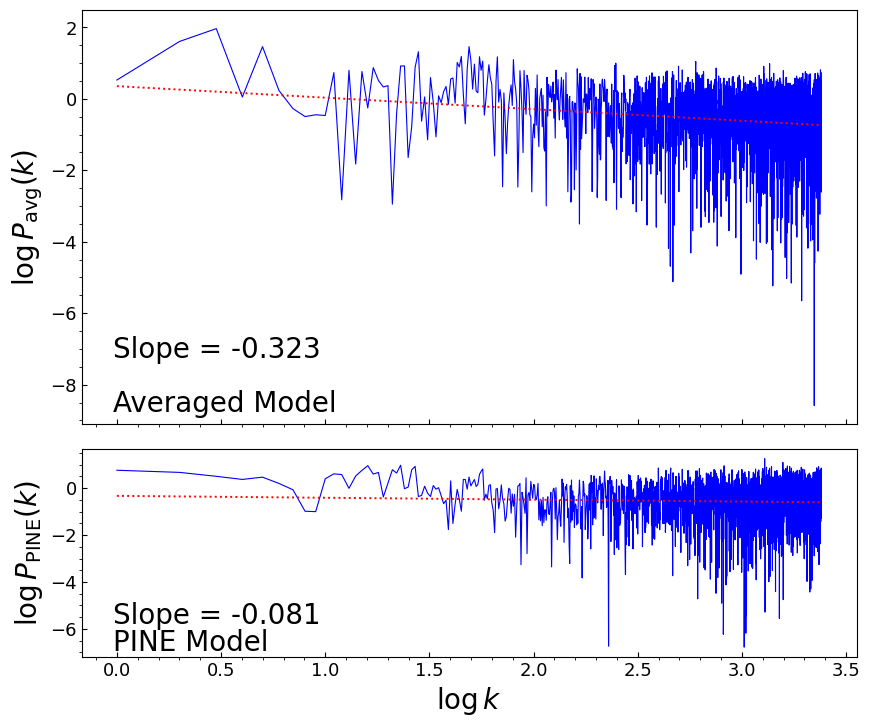}
\caption{
Fourier power spectra corresponding to the shell-based boustrophedon ordering.
}
\label{fig:pine_shell_fft}
\end{figure}

To investigate whether the remaining structures are correlated with shell effects, we use the shell-based boustrophedon ordering constructed from the distances to the nearest proton and neutron shell closures~\cite{BohrMottelsonI,BohrMottelsonII}, with the corresponding residual sequences and Fourier spectra plotted in Figs.~\ref{fig:pine_shell_residuals} and~\ref{fig:pine_shell_fft}, respectively.

The shell-based ordering reveals a yet more pronounced suppression of coherent structures. The averaged model retains a negative spectral slope of $\approx -0.323$, whereas the PINE spectrum approaches a slope of $\approx -0.081$. This behavior indicates that once the neural residual corrections are applied, the remaining fluctuations no longer exhibit strong shell-dependent long-range organization. Only weak residual low-frequency structures remain, primarily associated with isolated regions dominated by the FRDM component of the ensemble.

\subsubsection*{Two-dimensional Fourier analysis}

While the one-dimensional ordering characterizes global fluctuation behavior, it partially mixes proton and neutron correlations into a single sequence~\cite{PhysRevLett.94.102501,Hirsch2004-sl}. To separately investigate proton and neutron correlations, a direct two-dimensional Fourier transform was performed on the residual surface
\[
\Delta M(Z,N)
\]
following the general spectral-analysis framework previously employed in nuclear mass fluctuation studies~\cite{NIU2018759}.

\begin{figure}[H]
\centering
\includegraphics[width=0.98\columnwidth]{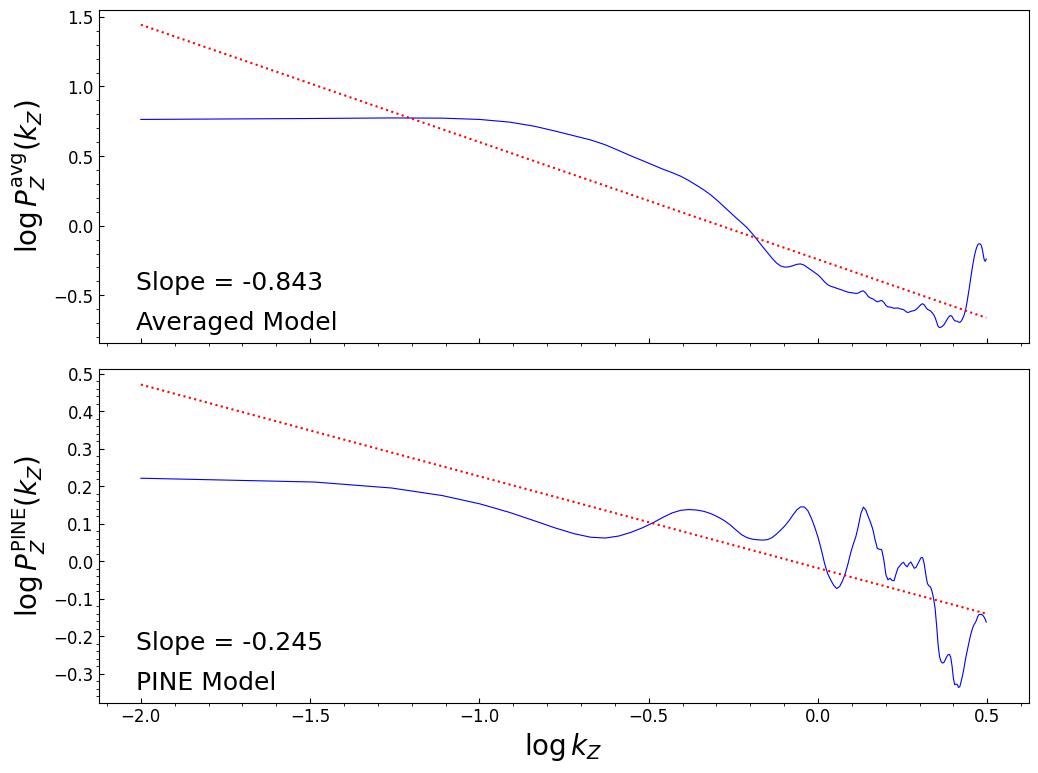}
\caption{
Projected proton-direction power spectrum $P_Z(k_Z)$.
}
\label{fig:PZ}
\end{figure}

\begin{figure}[H]
\centering
\includegraphics[width=0.98\columnwidth]{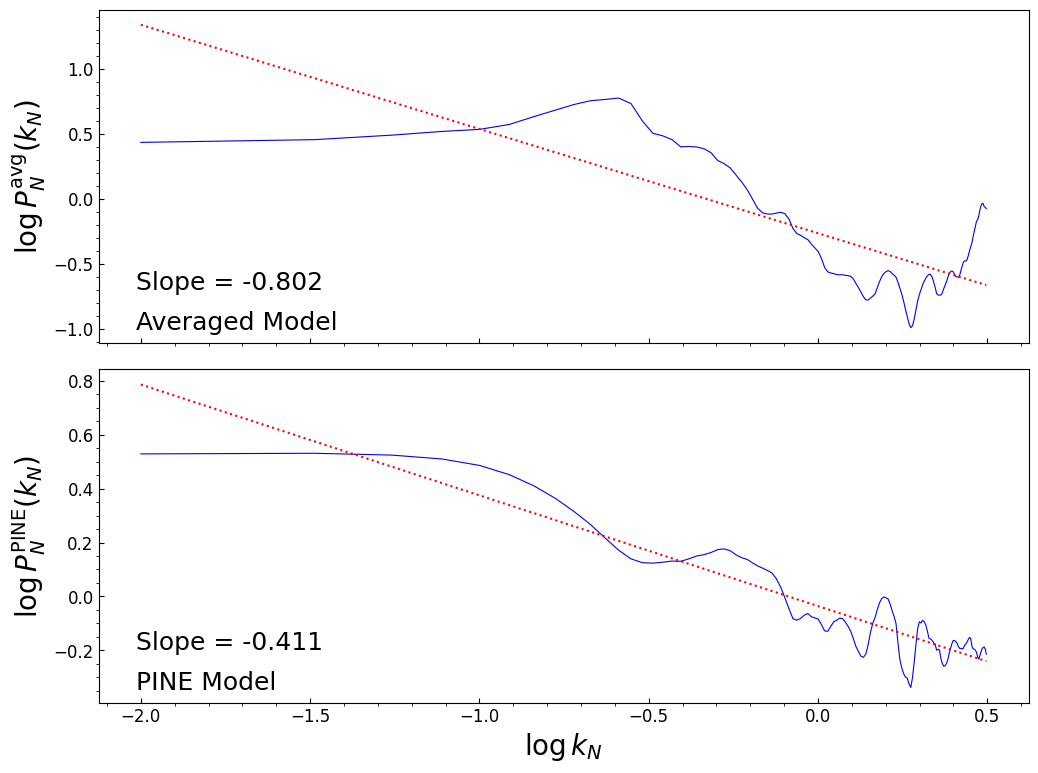}
\caption{
Projected neutron-direction power spectrum $P_N(k_N)$.
}
\label{fig:PN}
\end{figure}

The projected proton- and neutron-direction spectra shown in Figs.~\ref{fig:PZ} and~\ref{fig:PN} exhibit a strong suppression of low-frequency power after NN filtering. In the proton direction, the spectral slope decreases from about $-0.843$ for the averaged model to $-0.245$ for the PINE framework, while in the neutron direction it changes from roughly $-0.802$ to $-0.411$. Although weak low-frequency structures persist in the neutron projection, both spectra become noticeably flatter after filtering, indicating suppression of long-range correlations and survival of predominantly localized fluctuations. Similar sensitivity to local mass-systematics has also been reported in kernel-based and adaptive-learning approaches~\cite{Tian2025,Jalili2025}.
\FloatBarrier

\subsubsection*{Radial power spectrum}

\begin{figure}[H]
\centering
\includegraphics[width=0.98\columnwidth]{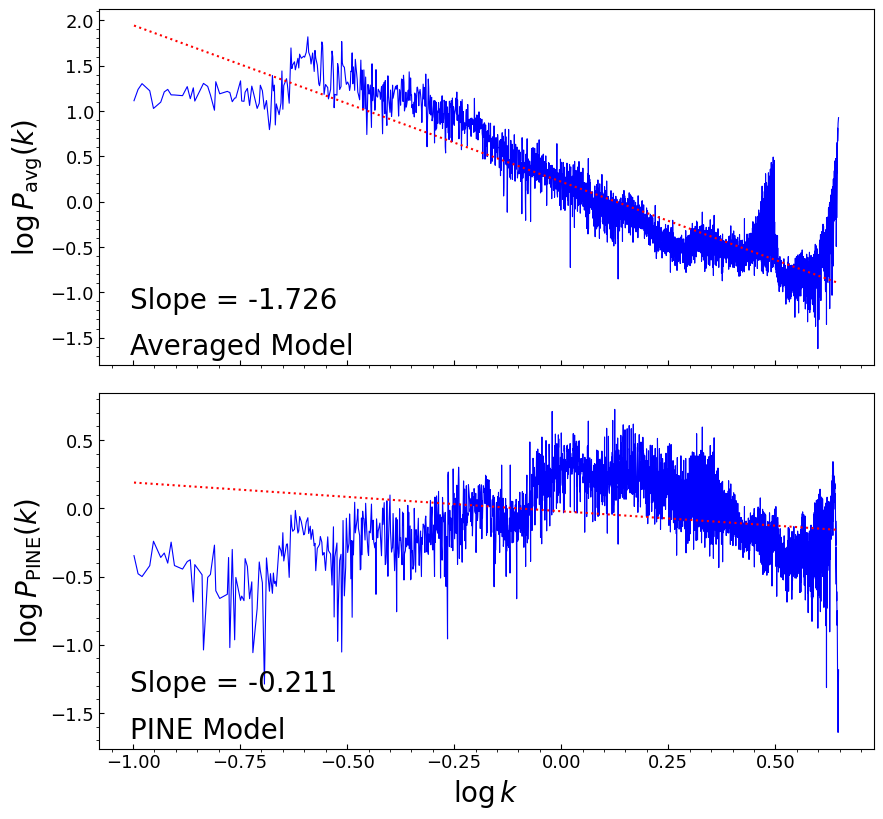}
\caption{
Radially averaged two-dimensional power spectrum.
}
\label{fig:Pradial}
\end{figure}

To study the isotropic fluctuation behavior, the two-dimensional power spectrum was also averaged radially, as shown in Fig.~\ref{fig:Pradial}. The radial spectrum shows one of the largest differences between the averaged model and the PINE residuals. The averaged model exhibits strong low-frequency enhancement with spectral slope $\approx -1.726$ (close to the chaotic $1/f^2$ limit), while the PINE spectrum becomes much flatter with slope $\approx -0.211$ (approaching the white-noise limit).

This behavior suggests that the HRD filtering stages successfully suppress the large-scale chaotic correlated structures not only along the proton and neutron directions separately, but also across the full two-dimensional residual field. By tuning the filtering capacity, the chaotic signature is systematically reduced, leaving behind isotropic, weakly correlated fluctuations.

\subsubsection*{Dyson--Mehta $\Delta_3$ statistics}
\begin{figure}
\centering
\includegraphics[width=0.98\columnwidth]{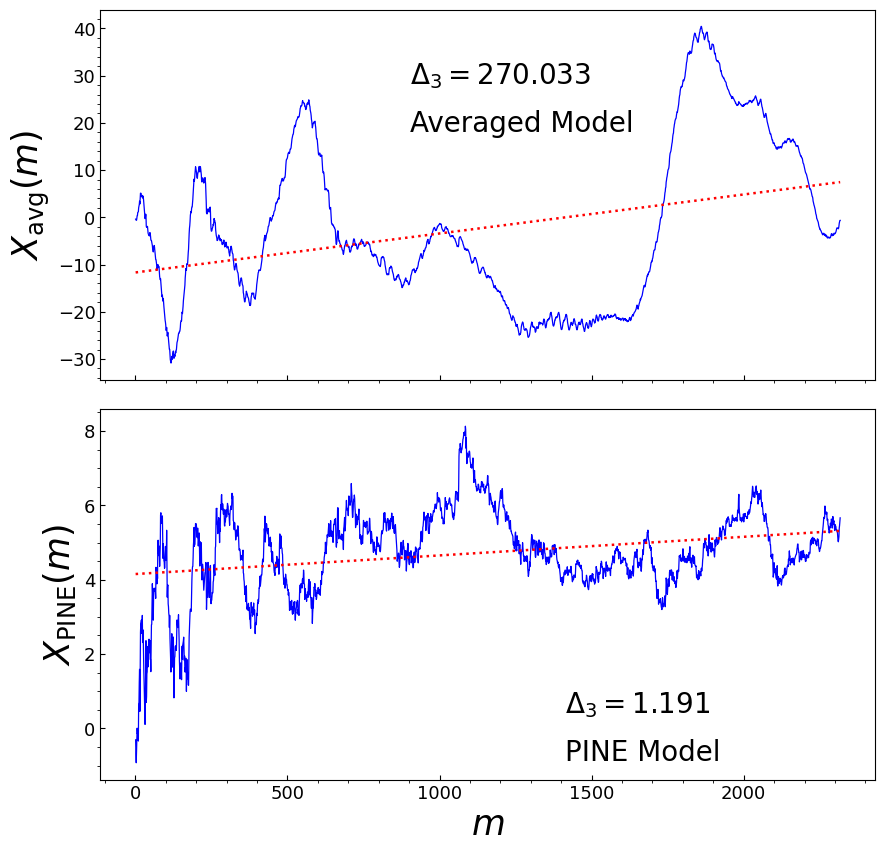}
\caption{
Dyson--Mehta $\Delta_3$ analysis for the averaged and PINE models.
}
\label{fig:delta3}
\end{figure}

To further investigate the long-range fluctuation behavior, the Dyson--Mehta $\Delta_3$ statistic was evaluated using the cumulative residual process, as illustrated in Fig.~\ref{fig:delta3}.
The averaged model exhibits extremely large long-range fluctuations,
\[
\Delta_3 \approx 270,
\]
reflecting strong, rigid collective correlations extending across large regions of the nuclear chart. After the PINE correction, the statistic decreases dramatically to
\[
\Delta_3 \approx 1.19,
\]
which represents a near-complete suppression of the spectral rigidity signature. This dramatic drop demonstrates that the hierarchical neural ensemble successfully eliminates the long-range correlations that are characteristic of quantum-chaotic systems, replacing them with a weakly correlated, Poisson-like noise distribution.

\begin{table}
\caption{
Parameters associated with the Dyson--Mehta $\Delta_3$ statistics.
}
\label{tab:delta3}
\begin{ruledtabular}
\begin{tabular}{cccc}
Model & $A$ & $B$ & $\Delta_3$ \\
\hline
Averaged Model & $0.008256$ & $-11.67856$ & $270.0332$ \\
PINE Model & $0.000501$ & $4.15252$ & $1.1906$ \\
\end{tabular}
\end{ruledtabular}
\end{table}

The one-dimensional Fourier spectra, shell-based orderings, two-dimensional projected spectra, radial spectra, and $\Delta_3$ statistics all show the same overall behavior. After the PINE filtering stages, the dominant low-frequency structures in the nuclear mass residuals are strongly suppressed, while the remaining fluctuations become much weaker and more localized.

\subsection{Inference from spectral analysis}
The spectral analysis shows that the dominant structures in the nuclear mass residuals mainly come from low-frequency correlations extending over broad regions of the nuclear chart~\cite{PhysRevLett.88.092502,Hirsch2004-sl,Hirsch2005-rr,PhysRevLett.96.042502,NIU2018759}. In the averaged model, the spectra show strong low-frequency enhancement together with large negative slopes, indicating that slowly varying correlated structures persist across neighboring nuclei. Such behavior is characteristic of residual fields that retain long-range correlated structures not fully absorbed by global averaged descriptions.

After the PINE correction, the spectra become much flatter in all the orderings and spectral projections studied in the present work. The largest changes happen mainly at low frequencies, where most of the original spectral enhancement disappears after the residual-learning stages. Since the HRD hierarchy is built recursively, the earlier NNs mainly remove the smoother large-scale structures. The later stages then act on weaker and more localized residual components. Because of this, the final spectra are no longer strongly dominated by long-range coherent behavior.

In the one-dimensional Fourier spectra, most of the spectral power in the averaged model is concentrated at low frequencies~\cite{PhysRevLett.94.102501,Hirsch2004-sl,Hirsch2005-rr,Relano2002,PhysRevLett.93.244101}. These low-frequency components are usually connected with broad systematic trends related to shell evolution, deformation, pairing, and other slowly varying collective contributions in the nuclear mass surface~\cite{BohrMottelsonI,BohrMottelsonII,Bender2003,FRDM2012,HFB24,DZ10}. After the PINE filtering stages, the low-frequency power becomes much smaller and the spectra move closer to flatter distributions.

The radial Fourier spectra show the same general behavior. In the averaged model, a large part of the spectral power is concentrated at small radial frequencies~\cite{NIU2018759}. In the PINE spectra, however, most of this low-frequency enhancement disappears and the remaining spectral power shifts more toward intermediate and shorter wavelengths. At the highest frequencies, only weak residual slopes remain, consistent with fluctuations that are local and short-range rather than organized across the full nuclear chart.

The redistribution of spectral power toward intermediate frequencies is also important. In correlated systems, the dominant spectral contribution is usually connected with slowly varying low-frequency modes~\cite{Relano2002,PhysRevLett.93.244101,PhysRevLett.118.204101}. After the NN filtering stages, much of this smooth long-range structure disappears, while weaker local fluctuations still remain. At the highest frequencies, the spectra still show some remaining structure instead of perfectly flat white-noise behavior. This suggests that part of the residual field retains short-range organization connected with local many-body effects~\cite{Aberg2002,MOLINARI200648,PhysRevLett.96.042502}.

The shell-based ordering also gives additional information about the origin of the remaining correlations. Since the nuclei are reorganized according to their distances from shell closures, shell-dependent structures become easier to see in the spectra~\cite{BohrMottelsonI,BohrMottelsonII,PhysRevLett.94.102501}. For the FRDM calculations, weak low-frequency components still remain after the HRD filtering stages. This suggests that part of the shell-related structure still survives in the residual field. These effects are less pronounced in the HFB and DZ calculations, where the post-filtered spectra become substantially flatter under the shell ordering~\cite{Bender2003,HFB24}. The remaining fluctuations in these cases are more localized and show weaker shell-dependent structure.

The projected two-dimensional spectra $P_Z(k_Z)$ and $P_N(k_N)$ also show a very similar behavior. In the averaged model, both the proton and neutron projections contain strong low-frequency components spread over large regions of the residual surface~\cite{NIU2018759,Hirsch2004-sl,PhysRevLett.94.102501}. After the PINE correction, the low-frequency enhancement becomes much smaller in both directions. The resulting spectra are considerably flatter, indicating that most slowly varying proton- and neutron-dependent structures are removed by the residual-learning hierarchy.

The radial spectra show the same general trend. The reduction of the low-frequency structure is not limited to only one direction in the $(Z,N)$ plane. Instead, the suppression happens more globally across the full residual surface. Most of the remaining spectral power is now concentrated at shorter wavelengths~\cite{Relano2002,PhysRevLett.93.244101,NIU2018759,Aberg2002}.

The Dyson--Mehta $\Delta_3$ statistic also supports the same picture~\cite{Dyson1962III,Mehta2004}. In the averaged model, the cumulative residual sequence shows strong long-range drift together with a very large $\Delta_3$ value. This is characteristic of rigid, long-range correlated behavior~\cite{Guhr1998,RevModPhys.81.539}. After the PINE correction, the cumulative fluctuations become much more localized and the $\Delta_3$ statistic decreases by almost two orders of magnitude. The remaining residual field is therefore much less rigid and moves closer to a weakly correlated regime~\cite{Relano2002,PhysRevLett.93.244101,PhysRevLett.96.042502}.

An important feature of the present results is that the post-filtered spectra look quite similar even when different mass models and ordering procedures are used. The DZ, FRDM, and HFB models all contain different systematic structures before filtering. After the HRD stages, however, the spectra from all of them converge toward a relatively narrow weakly correlated region. This suggests that most of the smooth large-scale structures that can be learned by the neural residual-learning procedure have already been removed. Most of the remaining fluctuations are now more local and connected with localized many-body effects~\cite{Aberg2002,PhysRevLett.96.042502,MOLINARI200648}.

A useful comparison can also be made with the Garvey--Kelson relations, which already produce nearly flat spectra even without using NNs~\cite{Garvey1966,Garvey1969,PhysRevLett.94.102501}. Owing to their strongly local structure, the Garvey--Kelson relations constrain the residual deviations through consistency conditions between neighboring nuclei~\cite{Garvey1966,Garvey1969}. As a result, large coherent low-frequency structures do not build up across the nuclear chart~\cite{Relano2002,PhysRevLett.93.244101}. The present global NN filtering procedure leads to a similar flattening of the spectra, although weak localized residual structures still remain after the HRD stages.

Overall, the Fourier spectra, shell-based orderings, projected two-dimensional spectra, radial spectra, and $\Delta_3$ statistics all show the same general behavior. The dominant low-frequency structures in the nuclear mass residuals become much weaker after the HRD and PINE filtering stages. Most of the remaining fluctuations are now weaker, more local, and less coherent across the nuclear chart~\cite{Relano2002,PhysRevLett.93.244101,PhysRevLett.118.204101,NIU2018759}. The surviving residual structures are concentrated predominantly in the lighter nuclei, while the medium- and heavy-mass regions exhibit considerably weaker long-range organization after filtering.

\section{Conclusions}

A multistage HRD framework based on FFNN and MoE residual learning has been developed and applied to the analysis of nuclear mass residuals. Rather than directly replacing the underlying physical mass models, the NN hierarchy operates as a controlled multi-scale filter that sequentially extracts progressively more localized structures from the residual surface. This construction systematically suppresses the dominant large-scale correlations and chaotic fluctuations, driving the remaining residuals toward the uncorrelated white-noise limit.

The resulting PINE model combines six independently trained residual learners constructed from the DZ10, FRDM, and HFB24 mass models together with both FFNN and MoE architectures~\cite{FRDM2012,HFB24,DZ10}. Through weighted hybridization of these complementary residual descriptions, the final framework substantially reduces the global prediction error while strongly suppressing coherent low-frequency spectral structures. The final residual distributions exhibit significantly flatter Fourier spectra and strongly reduced spectral rigidity, indicating that a large fraction of the long-range correlated behavior present in the original residuals has been systematically removed~\cite{Relano2002,PhysRevLett.93.244101,PhysRevLett.118.204101,Dyson1962III,Mehta2004}.

The spectral decomposition analysis demonstrates that the dominant structures remaining after the PINE correction are no longer associated with broad smooth collective trends extending across large regions of the nuclear chart. Instead, the surviving fluctuations become increasingly fragmented, localized, and concentrated at comparatively higher frequencies. This behavior indicates that the residual dynamics after the HRD filtering procedure are governed primarily by short-range many-body effects, local shell fluctuations, clustering phenomena, and rapid structural transitions rather than unresolved smooth global physics~\cite{Aberg2002,MOLINARI200648,PhysRevLett.96.042502}. The concentration of the remaining residual strength within restricted low-mass regions further supports the interpretation that the major coherent systematic structures accessible to global residual learning have already been systematically removed by the hierarchical decomposition procedure.

A key outcome of the present work is that the HRD framework functions not merely as a predictive NN correction scheme, but as a general spectral filtering methodology capable of isolating physical correlations. Because the sequential residual-learning architecture progressively separates structures according to their characteristic coherence scales, the framework naturally provides a decomposition of the residual dynamics into global, intermediate, and localized fluctuation components. Since the HRD framework operates directly on structured residual fields, the same approach can in principle be extended to a wide class of physical systems involving correlated fluctuations, including nuclear radii, decay properties, excitation spectra, and reaction observables. More broadly, the sequential residual-decomposition strategy extends naturally to machine-learning problems outside nuclear physics where the objective is not only predictive accuracy but also the systematic isolation of hidden correlation structures across multiple characteristic scales~\cite{Goodfellow2016,Bishop2006,Hastie2009,LeCun2015}.

\section*{Data and Code Availability}
The complete implementation of the HRD framework together with the PINE hybrid model developed in the present work is publicly available at Zenodo~\cite{Singh2026PINE}.

Experimental nuclear masses were taken from the AME2020 evaluation~\cite{AME2020,AME2020_II}.
Theoretical baseline models employed include FRDM(2012)~\cite{FRDM2012}, DZ10~\cite{DZ10}, and HFB-24~\cite{HFB24}.

\begin{acknowledgments}
C.Q. gratefully acknowledges the contribution of Fran\c{c}ois Vinet at the initial stage of this project.
\end{acknowledgments}

\bibliographystyle{apsrev4-2}
\bibliography{references}
\end{document}